 \documentclass[12pt,preprint]{aastex}

\usepackage{graphics}
\usepackage{graphicx}
\usepackage{times}
\usepackage{natbib}
\usepackage{amsmath}
\usepackage[english]{babel}
\usepackage{color}

\renewcommand{\d}{\mathrm{d}}
\newcommand{\gsim}{\ga}
\newcommand{\lsim}{\la}

\def\aap{A\&A}

\def\apjl{ApJ}

\def\apjs{ApJS}

\def\beq{\begin{equation}}
\def\eeq{\end{equation}}
\def\simlt{\mathrel{\rlap{\lower 3pt\hbox{$\sim$}}\raise 2.0pt\hbox{$<$}}}
\def\simgt{\mathrel{\rlap{\lower 3pt\hbox{$\sim$}} \raise 2.0pt\hbox{$>$}}}
\def\lsim{\mathrel{\rlap{\lower 3pt\hbox{$\sim$}}\raise 2.0pt\hbox{$<$}}}
\def\gsim{\mathrel{\rlap{\lower 3pt\hbox{$\sim$}} \raise 2.0pt\hbox{$>$}}}

\definecolor{Red}{rgb}{1, 0, 0}
\definecolor{Blue}{rgb}{0, 0, 1}
\definecolor{Black}{rgb}{0, 0, 0}
\definecolor{White}{rgb}{1, 1, 1}

\def\msun{\rm M_\odot}
\def\tsfr{t_\star^0}

\begin{document}


\title{Effects Of Circum-nuclear Disk Gas Evolution \\ And The Spin Of Central Black Holes}

\author{Umberto~Maio\altaffilmark{1}}
\affil{{}$^1~$ Max Planck Institute for Extraterrestrial Physics, Giessenbachstra{\ss}e, 85748 Garching b. M\"unchen, Germany; Osservatorio Astronomico di Trieste, via G.~B. Tiepolo 11, 34143 Trieste, Italy; Leibniz Institute for Astrophysics, An der Sternwarte 16, 14482 Potsdam, Germany}
\author{Massimo Dotti\altaffilmark{2,3}}
\affil{{}$^{2,3}~$Department of Physics of the University of Milano-Bicocca, Piazza della Scienza 3, 20126, Milano, Italy;INFN, Sezione di Milano-Bicocca, Piazza della Scienza 3, 20126 Milano, Italy}
\author{Margarita Petkova\altaffilmark{4,5}}
\affil{{}$^{4,5}~$Max Planck Institute for Astrophysics, Karl-Schwarzschild-Stra{\ss}e 1, 85741 Garching b. M\"unchen, Germany; Department of Astronomy, University of Bologna, via Ranzani 1, 40127, Bologna, Italy}
\author{Albino Perego\altaffilmark{6}}
\affil{{}$^6~$Department of Physics, University of Basel, Klingelbergstra{\ss}e 82, 4056, Basel, Switzerland}
\author{Marta Volonteri\altaffilmark{7,8}}
\affil{{}$^{7,8}~$Institut d'Astrophysique de Paris, 98bis Boulevard Arago, 75014 Paris, France; Department of Astronomy, University of Michigan, Ann Arbor, Michigan 48109, USA}
%


\begin{abstract}
\noindent
Mass and spin are the only two parameters needed to completely characterize black holes in General Relativity. However, the interaction between black holes and their environment is where complexity lies, as the relevant physical processes occur over a large range of scales. That is particularly relevant in the case of super-massive black holes (SMBHs), hosted in galaxy centers, and surrounded by swirling gas and various generations of stars. These compete with the SMBH for gas consumption and affect both dynamics and thermodynamics of the gas itself.
How the behavior of such fiery environment influence the angular momentum of the gas accreted onto SMBHs, and, hence, black-hole spins is uncertain. We explore the interaction between SMBHs and their environment via first 3D sub-parsec resolution simulations (ranging from $\sim 0.1~\rm pc$ to $\sim 1~\rm kpc$ scales)  that study the evolution of the SMBH spin by including the effects of star formation, stellar feedback, radiative transfer, and metal pollution according to the proper stellar yields and lifetimes.
This approach is crucial to investigate the impact of star formation processes and feedback effects on the angular momentum of the material that could accrete on the central hole.
We find that star formation and feedback mechanisms can locally inject significant amounts of entropy in the surrounding medium, and impact on the inflow inclination angles and Eddington fractions.
As a consequence, the resulting trends show upper-intermediate equilibrium values for the spin parameter of $a\simeq 0.6-0.9$, 
corresponding to radiative efficiencies $\epsilon \simeq 9\%-15\%$.
These results suggest that star formation feedback taking place in the circum-nuclear disk during the infall cannot induce alone very strong chaotic trends in the gas flow, quite independently from the different numerical parameters.
\end{abstract}


\object{Accepted -- referee version}



\section{Introduction}\label{sect:introduction}

\begin{figure}
\centering
 \includegraphics[width=0.4\textwidth, height=0.18\textheight]{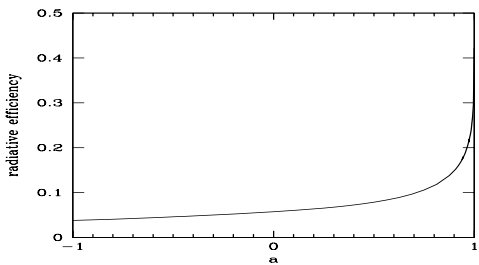}
 \caption[Radiative efficiency]{\small
Radiative efficiency as a function of the spin parameter, $a$ \cite[][]{Bardeenetal1972, NovikovThorne1973}. Positive values of $a$ correspond to SMBHs co-rotating with their accretion disks, while negative values for $a$ indicate counter-rotating SMBHs.
}
\label{fig:epsilona}
\end{figure}
\begin{figure}
\centering
\includegraphics[width=0.45\textwidth]{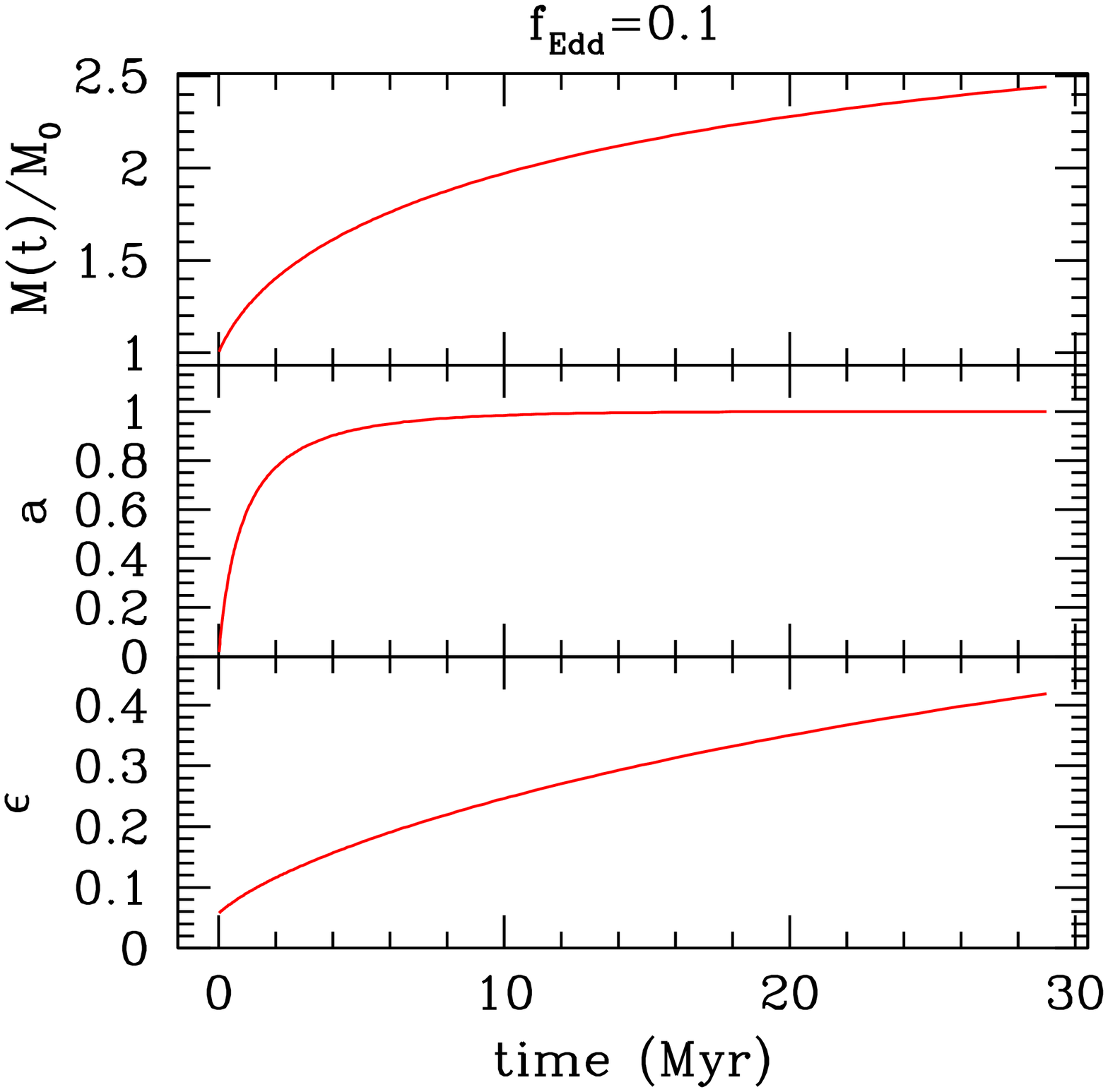}
  \caption[Radiative efficiency]{\small
	Time evolution of the properties of a black hole with initial mass $M(0)$, when accreting from a prograde accretion disk at a rate corresponding to $10\%$ the Eddington ratio.
	Top panel: relative mass increase, $M(t)/M(0)$.
	Middle panel: spin parameter evolution, $a$, derived according to \cite{Bardeen1970}'s formalism.
	Bottom panel: corresponding evolution of the radiative efficiency, $\epsilon$.
   }
\label{fig:efficiency}
\end{figure}

\noindent
Black holes (BH) in Einstein's theory of general relativity are described by only three parameters: mass, spin and charge.
Astrophysical black holes are even simpler, as charge can be neglected.  So, besides their masses, $M_{\rm BH}$, they are completely characterized by their dimensionless spin parameter, $a \equiv c \, J_{BH}/G \, M_{\rm BH}^2$, where 
$c$ is the speed of light,
$G$ the gravitation constant,
$J_{\rm BH}$ is the angular momentum of the black hole, and 
$0\le a\lesssim 1$ \cite[e.g.][]{Kerr1963,Peters1964,BoyerLindquist1967,Carter1968,Felice1968,Bardeenetal1972}.
\\
Many previous theoretical efforts have focused on the mass growth of SMBHs that reside in galaxy centers and power quasars \cite[e.g.][and references therein]{Lynden-Bell1969,Bardeen1970,ShakuraSunyaev1973,ShakuraSunyaev1976,Blandford1977,Abramowicz1978,Rees1982,WilsonColbert1995,ModerskiSikora1996,GhoshAbramowicz1997,Moderski1998,Silk1998,HNR1998,Livio1999,DiMatteo2005,Dimatteo2008,Bower2006,Croton2006,Li2006,King2006,KingPringle2007,Sijacki2007,Booth2009,Sijacki2009,hopkins2005,Hopkins2011,hopkinsquataert2010,Fabian2010,Dubois2012}.
Thus, we will focus here on the processes taking place in their environments and on the consequent implications for the other property of SMBHs: spin, which is of no less importance.
\\
Indeed, the spin of a black hole affects the efficiency \cite[e.g.][and references therein]{NovikovThorne1973,BambiBarausse2011} of `classical' accretion processes.
In radiatively efficient thin accretion disks the mass-to-energy conversion efficiency, $\epsilon$, equals \cite[][]{NovikovThorne1973} $\epsilon\equiv 1- E/c^2$, where $E$ is the binding energy per unit mass of a particle at the last stable orbit.
The closer the last stable orbit is to the horizon, the higher the mass-to-energy conversion efficiency, which increases from $\epsilon\simeq 5.7\%$, for a non-rotating ($a=0$) hole, to $\epsilon\simeq 42\%$, for its maximally rotating counterpart 
(see figure~\ref{fig:epsilona}).
\\
The mass-to-energy conversion directly affects the mass growth rate of black holes \cite[][]{Soltan1982,YuTremaine2002,Shankar2009} according to
\begin{equation}
  \epsilon~\frac{\d M}{\d t} = f_{\rm Edd}~\frac{L_{\rm Edd}}{c^2}
\end{equation}
where $ f_{\rm Edd} $ is the Eddington efficiency, and $L_{\rm Edd}$ the Eddington luminosity:
high radiative efficiency, $\epsilon$, implies slow growth, as more mass is radiated away instead of being fed into the black hole.
For instance, for a black hole accreting at a fraction $f_{\rm Edd}$
of the Eddington rate, $L_{\rm Edd}$, assuming constant $\epsilon$, the
mass increases with time as:
\begin{equation} \label{eq:Mt}
  M(t)=M(0)\,\exp\!\left\{{f_{\rm Edd} {(1-\epsilon)}\over{\epsilon}}\frac{t}{t_{\rm Edd}}\right\},
\end{equation}
where $M(0)$ is the initial mass, and the Eddington timescale, $t_{\rm Edd}$, is $c^2$ divided by the Eddington luminosity per unit mass,
\begin{equation}
  t_{\rm Edd} = \frac{ \sigma_T c}{4 \pi G m_p} \simeq 0.45\,{\rm Gyr},
\end{equation}
with
$\sigma_T$ Thomson cross section, and $m_p$ proton mass.
Moreover, the higher the spin parameter, $a$, the higher $\epsilon$, implying longer timescales to grow the black-hole mass (see also later figure~\ref{fig:efficiency}).
Going from $\epsilon\simeq 5.7\%$ to $\epsilon\simeq 42\%$, the difference in mass amounts to 6 orders of magnitude at $t=t_{\rm Edd}$, for $f_{Edd} = 1$. 
The typical spin therefore impacts the overall mass growth of SMBHs.
This is particularly important for the case of SMBHs powering the highest-redshift ($z\gsim 6$) quasars \cite[][]{KingPringle2007}, since these SMBHs must grow in mass by gas accretion in less than a billion year.
\\
\noindent
Although the initial mass of the SMBH plays a crucial role in this game, all the SMBH formation models proposed to date require significant and fast mass accretion to result in $10^9 \msun$ SMBHs at redshift $\gsim 6$. As an example, under the assumption of a constant spin and radiative efficient accretion, a non-rotating ($a=0$, $\epsilon \simeq 5.7\%$) SMBH accreting at its Eddington limit ($f_{\rm Edd} = 1$) can grow its mass of 6 orders of magnitude in only $ < 0.4\,\rm Gyr$.
A maximally rotating ($a=1$, $\epsilon \simeq 42\%$) SMBH would accrete (at $f_{\rm Edd} = 1$) the same mass in more than a billion year.
Note however that assuming a constant $a=0$ is highly unphysical, since non rotating SMBHs are spun-up by every accretion event with finite angular momentum.
Fortunately, the constraint on the spin parameter are not too stringent, since 
(see figure~\ref{fig:epsilona})
the radiative efficiency is a very steep function of $a$ only for $a \gtrsim  0.9$ \citep{NovikovThorne1973, KingPringleHofmann2008}, with typical values varying from a few per cents up to $\epsilon \approx 10\%$ for $a \lesssim 0.7$, and slightly increasing to $\epsilon \approx 15\%$ for $a \approx 0.9$.
As an example, from equation~(\ref{eq:Mt}) one can estimate that a SMBH with $a \approx 0.8$ ($\epsilon \approx 11.5\%$) would grow by 6 orders of magnitude in roughly two times the time needed by a non-rotating SMBH still well within the age of the Universe at $z\approx 6$ (of about $ 1~\,\rm Gyr$).
\\
The magnitude and orientation of SMBH spins also affect the frequency of SMBHs in galaxies, via `gravitational recoil'.
When two black holes coalesce, they may recoil due to anisotropic emission of gravitational waves. If this recoil were sufficiently violent, the newly merged hole would escape from the host galaxy. Recent breakthroughs in numerical relativity have allowed reliable
computations of recoil velocities. Non-spinning black holes are
expected to recoil with velocities below 200 $\rm{km\,s^{-1}}$. The recoil is much larger, up to thousands $\rm{km\,s^{-1}}$, for SMBHs with large spins in non-aligned configurations 
\cite[e.g.][]{Campanelli2007, Lousto2012}.
\\
Finally, the value of the spin parameter, $a$, in a Kerr black hole \cite[][]{Kerr1963, BoyerLindquist1967} also determines how much energy is in principle extractable from the black hole itself \cite[][]{Blandford1977, BlandfordPayne1982}, and it may be important \cite[as argued by e.g.][]{MacDonaldThorne1982, ModerskiSikora1996, GhoshAbramowicz1997, McKinney2005} in relation to jet production \cite[but see][]{Livio1999}.
\\
In general, since SMBH growth is supposed to be led by mass accretion \cite[][]{Soltan1982,YuTremaine2002,Shankar2009,Shankar2011arXiv}, it is the way SMBHs accrete gas that will determine their spin.
Spin-up is a natural consequence of prolonged disk-mode accretion: 
any hole that has, for instance, doubled its mass
by capturing material with constant angular momentum axis (coherent
accretion) would end up spinning rapidly, close to the maximum allowed value \cite[][]{Thorne1974}.
For example, if a BH accretes at values of order $\sim 10\%$ the Eddington limit (i.e. $f_{\rm Edd} = 0.1$), then  it can be spun up to high values in less than 10~Myr (figure~\ref{fig:efficiency}).
The radiative efficiency evolves much more slowly, since as discussed previously, it is a shallow function of the spin until $a\sim 0.9$.
\\
On the other hand, a black hole which had gained its mass from capturing many low-mass objects in randomly-oriented orbits or from small (and short) uncorrelated accretion episodes (chaotic accretion), would keep small spin \citep[$a<0.2$, e.g.][]{King2006,KingPringle2007}.
\\
The main unknown here is therefore the typical angular momentum of the material feeding the SMBH: does the vector maintain a roughly constant preferential direction, or does the direction change chaotically?
\\
Since SMBHs are expected to have grown in mass mostly during active phases as quasars, the evolution of SMBH spins has to be addressed by simulating the typical environmental conditions expected around quasars, where the properties of the accretion flow are established.
When accretion is triggered by galaxy mergers, as expected for quasars \cite[][]{SandersMirabel1996, Downes1998, Davies2003, Davies2004, DownesEckart2007}, the material that feeds a SMBH is expected to assemble into a circum-nuclear disk (CND), as also found by numerical simulations \cite[e.g.][]{Mayer2007}.
These are therefore the most relevant circumstances to be explored.
CNDs are also associated to very high levels of star formation that are expected to influence SMBH spin evolution: local star formation injects energy and momentum into the gas near a SMBH, possibly breaking the coherency of the gas inflow.
A quasar could be fueled by a sequence of well-separated short-lived events with randomly distributed angular momentum vectors, leading to a chaotic gas flow \cite[][]{King2008}.
\\
In this work we will try to understand if a coherent circum-nuclear disk around a SMBH can be affected by the gaseous and stellar processes taking place during infall.
Moreover, we will try to address if these phenomena can break or alter the coherency of the matter flow, independently from larger-scale effects, like mergers or SMBH-SMBH interactions
\cite[as already studied by e.g.][]{Hopkins2011}.
\\
Therefore, we simulate the gaseous and stellar environment around a $4 \times 10^6\,\msun $ SMBH placed in a circum-nuclear disk of $10^8\, \msun$, by using the parallel tree/SPH code Gadget-3, an improved version of the publicly available code Gadget-2 \cite[][]{gadget,Springel2010}, including cooling
\cite[][]{Maio2007,Maio2010,Maio2011}, star formation
\cite[][]{SpringelHernquist2003}, stellar, and radiative feedback
\cite[][]{Petkova2011}.
We make sure that our spatial resolution (from 0.1 to 1~pc) is sufficient to resolve the `sphere of influence' within which the SMBH dominates gravity ($\sim$ 10-20~pc), allowing us to study the dynamics of the inflowing material \cite[e.g.][]{Dotti2010,Hopkins2011}.
This also allows us to model gas cooling, star formation, and stellar feedback accurately.
In our investigation we focus on the effect of local perturbations, neglecting large-scale instabilities that can alter the angular momentum of the nuclear gas \cite[e.g.][]{Hopkins2011}.
\\
The paper is organized as follows: in section~\ref{sect:simulations}, we introduce and describe the simulations we performed; in section~\ref{sect:spin} and \ref{sect:spin results}, we outline the technique used to address spin evolution and show our main results; in section \ref{sect:conclusions}, we comment and conclude.


\section{Simulations}\label{sect:simulations}


\noindent
In the following we will describe the code used, the physical processes included in our treatment, and the set of 3D simulations performed.


\subsection{Implementation}\label{sect:implementation}

\noindent
We run 3D N-body/hydro simulations by using the parallel tree, SPH code Gadget3, an extended version of the publicly available code Gadget2 \cite[][]{Springel2005,Springel2010}.  Besides gravity and hydrodynamics, the code includes gas cooling down to low temperatures \cite[][]{Maio2007,Maio2010} and a multi-phase model for star formation \cite[][]{Springel2003,Hernquist2003}, inspired by the works of \cite{KatzGunn1991, Cen1992, CenOstriker1992, Katz1992, Katz_et_al_1992, Katz_et_al_1996}.
Additionally, mechanical feedback \cite[][]{Aguirre_et_al_2001} from supernovae (SN) is taken into account \cite[][]{Springel2003}.
We also implement radiative transfer (RT) prescriptions \cite[][]{Petkova2009,Petkova2011,PetkovaMaio2011} to model the
effects onto the surrounding medium of photo-ionization and photo-heating from stellar radiation and gaseous radiative processes.
\\
The main agents of gas cooling are atomic resonant transitions that can rapidly bring the temperatures down to $\sim 10^4\,\rm K$. In metal-rich environments, atomic fine-structure transitions can lower temperatures further to $\sim$ 10 K \cite[][]{Maio2007,Maio2010b}. We will additionally show that detailed metallicity evolution can slightly influence the resulting star formation and feedback processes, but has no drastic effects on the spin evolution of the central SMBH and helps stabilizing the disk against star-formation-induced chaotic motions.
\\
When the medium cools and gets denser than a given density threshold of $10^4$ cm$^{-3}$, stochastic star formation is assumed, cold gas is gradually converted into stars, and entropy feedback is injected into the surrounding environment, leading to a self-regulated star forming regime. The only free parameter in this model is the maximum star formation timescale, $\tsfr$, which determines the amplitude of the typical star formation timescale, $t_\star$, at any gas density, $\rho$, above the mass-density threshold, $\rho_{\rm th}$,
\begin{equation}
  \label{tsfr}
  t_\star(\rho) = \tsfr \left( \frac{\rho}{\rho_{\rm th}} \right)^{-1/2}.
\end{equation}
In some astrophysical problems, it is possible to fix $\tsfr$ by matching with observations\footnote{
For example, in galactic disks, $\tsfr=2.1\,\rm Gyr$ allows us to fit the Kennicutt-Schmidt law \cite[][]{Kennicutt1998,Springel2003}.
}.
In strongly dynamically evolving systems, such as intense star-bursts in galactic nuclear regions, however, it is impossible to define a unique characteristic timescale for star formation.
Thus, we will assume two extreme timescales, a very long one ($\tsfr=300\,\rm Myr$), and a very short one  ($\tsfr=1\,\rm Myr$), as limiting cases \cite[e.g.][]{Krumholz2011}.
This means that the maximum timescales at the threshold for star formation are $\rm 300\,\rm Myr$ and $1\,\rm Myr$, respectively, while at $\rho > \rho_{th}$, the time over which stars can form, $t_\star(\rho)$, becomes shorter and shorter.
Moreover, we will also explore the implications of intermediate, more realistic values ($\tsfr=6, 10, 20, 50, 100\,\rm Myr$) in terms of star formation efficiency, and the effects of full stellar evolution with consistent metal enrichment calculations.
\\
At each timestep, $\Delta t \ll 1\,\rm Myr$, collisionles star particles are spawned with a probability distribution
\begin{equation}\label{probability_stars}
  p_{\star} = \frac{m}{m_\star} \left[ 1 - e^{-(1-\beta)x\Delta t / t_\star} \right],
\end{equation}
where $m$ is the current mass of the gas particle, $m_\star$ is the mass of each star particle, $\beta\simeq 0.1$ is the
fraction\footnote{Here a Salpeter IMF has been assumed over the range [0.1,100] M$_\odot$.
}
of short-lived stars which die instantly as supernov{\ae}, and $x$ is the cold-cloud mass fraction, ranging roughly from $0.85$ up to $\sim 1$ \cite[][]{Kennicutt1998, Springel2003}.
The star mass is given by $m_\star=m_0/N_g$, with $N_g$ an integer number indicating how many stars each gas particle can generate -- 
usually of the order of unity or a few \cite[][]{Springel2003, Hernquist2003}) -- and $m_0$ the initial gas particle mass, which is decreased (and reduced to the current gas mass $m$) accordingly to the spawned stars.
The resulting star formation rate is simply given by 
$\dot M_\star = (1-\beta)xm / t_\star$.
The initial position and velocity of each star particle are
assumed to be the same as the parent gas particle.
We note that relations (\ref{tsfr}) and (\ref{probability_stars}) imply higher probability of forming stars at higher densities.
Once the star forming regime kicks in, supernova energy heats the 
ambient medium and cold clouds are evaporated by thermal conduction: this process is parameterized via the evaporation factor, $A_0$, dependent on the environment.
Assuming that thermal instabilities in the gas become active at 
$\sim 10^5\,\rm K$ and typical SN temperatures are $T_{SN}\sim 10^8\,\rm K$, the evaporation factor is $A_0=T_{SN}/10^5\rm K\sim 10^3$.
However, different SN and cooling features would give different values, thus, we will assume $A_0 = 10^2$, $10^4$, and $10^6$.
For further details see \cite{Springel2003}.
Similarly, outflows are phenomenologically taken into account according to the following probability distribution:
\begin{equation}
  \label{probability_winds}
  p_{w} = 1 - e^{-\eta(1-\beta)x\Delta t / t_\star},
\end{equation}
where the wind efficiency, $\eta\sim 2$ \cite[][]{Martin1999,Springel2003},
is the ratio between the mass-loss rate and the star formation rate of each gas particle.
Winds carry an additional kinetic energy that is a fraction, $\chi\simeq 0.25$ \cite[][]{Springel2003}, of the supernova energy.
Therefore, the wind velocity, $v_w$, is $ v_{w} = \sqrt{2\chi\varepsilon_{\rm SN}/\eta \phantom{i}}$, where $\varepsilon_{\rm SN}=4 \times 10^{48}$ erg $\msun^{-1}$ is the supernova energy injected in the medium for each solar mass in stars formed \cite[assuming a Salpeter IMF,][]{Springel2003}.
The corresponding wind velocity has typical magnitudes of a few hundreds km/s in a direction that can be randomly chosen either on the unit sphere (isotropic winds) or along the axis uniquely defined by ${\bf v}~\times~{\bf\nabla}~\phi$, with velocity vector ${\bf v}$ and gravitational potential $\phi$ (axial winds).
\\
The photo-ionization and photo-heating of the gas is followed by a radiative transfer scheme extended to a multi-frequency regime, with four frequency bins in the range 13.6 -- 60~eV, and corresponding to the ionization potential of neutral hydrogen (13.6~eV), neutral helium (24.6~eV), singly ionized helium (54.4~eV), and one arbitrarily higher value (60~eV). The radiative transfer equations are coupled to hydrodynamics and solved self-consistently taking account both gas cooling and radiative heating from stellar sources \cite[for further details see][]{Petkova2009,Petkova2011,PetkovaMaio2011}. Ionizing radiation from stars and gas is diffused throughout the simulation volume, interacting with both hydrogen and helium. The stellar emissivity is assumed to have a black-body spectrum with effective temperature $T_{\rm eff} = 5\times 10^4 \, \rm K$ (usual emission from short-lived massive stars). The gas emissivity due to cooling processes is also a black-body spectrum with effective temperature equal to its own temperature.
\\
A few studies \cite[][]{wada2009,schartmann2009,schartmann2011} attempted analyses of the gas properties around SMBHs. None of them, though, was suited to study how star formation influences the evolution of SMBH spins, which is the main goal of our investigation. Previous works, in fact, made several simplifications on the physical processes considered, for instance neglecting either radiative transfer effects or the fundamental link between stellar evolution and the gaseous environment where stars form. Our simulations, instead, follow star formation and stellar evolution explicitly, so that SN explosions are driven from the hydrodynamical evolution of the gas. The ensuing feedback effects are therefore fully self-consistent with the dynamics and thermodynamics of the circum-nuclear disk.  These features allow us to investigate accurately how stellar and SN feedback affect the angular momentum of the gas near SMBHs, and, as a consequence, how star-bursts impact SMBH spins.


\subsection{Initial conditions}\label{sect:ICs}

\noindent
We initialize every simulation with a SMBH of $4\times 10^6 \msun$ at rest, at the center of a purely gaseous CND. The SMBH is treated as an active evolving collisionles particle, and reacts to the gravitational torques exerted by the stellar and gaseous background.
\\ 
The SMBH and the disk are embedded in a spherical stellar bulge \cite[][]{Magorrian1998} with total mass of $2\times 10^9 \msun$ and scale length $b=10^3\,\rm pc$, always modelled using $10^5$ particles of $2\times 10^4\msun$. The bulge follows a Hernquist mass density profile \cite[][]{Hernquist1990,Tremaine_et_al_1994}, which reproduces the $R^{1/4}$-law well \cite[][]{deVaucouleurs1948} and mimics the potential of the host galaxy.
\\
The disk is assumed to have a mass of $10^8 \msun$ (25 times larger than the central SMBH) and is modelled using $2 \times 10^4$, $2\times 10^5$, $2\times 10^6$, or $2 \times 10^7$ gas particles in our low, standard, high, and very-high resolution runs, such that the corresponding gas particle masses are $5000\,\msun$, $500\,\msun$, $50\,\msun$, and $5\,\msun$.
The CND is initially composed only of gas particles, since we aim at modelling the evolution of SMBHs in a recently formed CND. We note, however, that the CND becomes populated of young stars as the simulation evolves. The absence of an old population of stars in our initial conditions does not affect our results: only young stars, through their powerful winds or their deaths as supernovae, can significantly alter the properties of the gas.
The disk initially follows an exponential surface density profile, and has a scale length of $100\,\rm~pc$. It is pressure supported in the vertical direction, with an aspect ratio $H/R \equiv 0.1$. The disk internal energy is initially fixed according to a polytropic equation of state with adiabatic index $\gamma=5/3$ and cosmic composition for H and He.
\\
We evolve adiabatically our initial conditions for $\approx 100\,\rm Myr$, to ensure that the system reaches a stable configuration without fragmenting nor developing any strong asymmetry and instability, before turning on cooling, star formation, winds, supernova and radiative feedback.


\subsection{Runs}\label{sect:runs}

\noindent
\begin{table*}
  \centering
      {\footnotesize 
        \caption[Simulation set-up]{\small
          Initial parameters for the different runs: the black-hole mass is always fixed to $4\times 10^6\rm\msun$, the disk gas mass to $10^8\msun$, and the stellar bulge mass to $2\times 10^9\msun$.
        }
        \begin{tabular}{lccccccc}
          \hline
          \hline
          Runs & Particle number & Particle mass [$\rm \msun$]& Evap.  & SF time& Low-T & RT &Softening\\
          & gas $\quad$ (bulge) & gas $\quad$ (bulge) &  factor & [Myr] &cooling & & length [pc]\\
          \hline
          EVP1e2-SFRT1 	& $2\times 10^5\quad$ ($10^5$) & $500\quad$ ($2\times 10^4$) & $10^2$ & 1   & off &off &  1.0 \\
          EVP1e2-SFRT1-LowT$^{*}$ 	& $2\times 10^5\quad$ ($10^5$) & $500\quad$ ($2\times 10^4$) & $10^2$ & 1    & on  &off &  1.0 \\
          EVP1e2-SFRT300 	& $2\times 10^5\quad$ ($10^5$) & $500\quad$ ($2\times 10^4$) & $10^2$ & 300  & off &off &  1.0 \\
          EVP1e2-SFRT300-LowT$^{*}$ & $2\times 10^5\quad$ ($10^5$) & $500\quad$ ($2\times 10^4$) & $10^2$ & 300 & on  &off &  1.0 \\
          \hline
          EVP1e4-SFRT1 	& $2\times 10^5\quad$ ($10^5$) & $500\quad$ ($2\times 10^4$) & $10^4$ & 1   & off &off &  1.0 \\
          EVP1e4-SFRT1-LowT$^{*}$ 	& $2\times 10^5\quad$ ($10^5$) & $500\quad$ ($2\times 10^4$) & $10^4$ & 1    & on  &off &  1.0 \\
          EVP1e4-SFRT300 	& $2\times 10^5\quad$ ($10^5$) & $500\quad$ ($2\times 10^4$) & $10^4$ & 300  & off &off &  1.0 \\
          EVP1e4-SFRT300-LowT$^{*}$ & $2\times 10^5\quad$ ($10^5$) & $500\quad$ ($2\times 10^4$) & $10^4$ & 300  & on  &off &  1.0\\
          \hline
          EVP1e6-SFRT1 	& $2\times 10^5\quad$ ($10^5$) & $500\quad$ ($2\times 10^4$) & $10^6$ & 1   & off &off &  1.0 \\
          EVP1e6-SFRT1-LowT$^{*}$ 	& $2\times 10^5\quad$ ($10^5$) & $500\quad$ ($2\times 10^4$) & $10^6$ & 1    & on &off &  1.0 \\
          EVP1e6-SFRT300 	& $2\times 10^5\quad$ ($10^5$) & $500\quad$ ($2\times 10^4$) & $10^6$ & 300  & off &off &  1.0 \\
          EVP1e6-SFRT300-LowT$^{*}$ & $2\times 10^5\quad$ ($10^5$) & $500\quad$ ($2\times 10^4$) & $10^6$ & 300  & on  &off &  1.0 \\
          \hline
          EVP1e2-SFRT1-IW$^{\diamondsuit}$ & $2\times 10^5\quad$ ($10^5$) & $500\quad$ ($2\times 10^4$) & $10^2$ & 1   & off &off &  1.0 \\
          EVP1e2-SFRT300-IW$^{\diamondsuit}$ & $2\times 10^5\quad$ ($10^5$) & $500\quad$ ($2\times 10^4$) & $10^2$ & 300   & off &off &  1.0 \\
          \hline
          EVP1e2-SFRT1-LR$^\dag$ 	& $2\times 10^4\quad$ ($10^5$) & $5000\quad$ ($2\times 10^4$) & $10^2$ & 1   & off &off &  1.0 \\
          EVP1e2-SFRT1-HR$^\star$ 	& $2\times 10^6\quad$ ($10^5$) & $50 \qquad$ ($2\times 10^4$) & $10^2$ & 1   & off &off &  1.0 \\
          EVP1e2-SFRT1-HR-SS$^\star$ & $2\times 10^6\quad$ ($10^5$) & $50 \qquad$ ($2\times 10^4$) & $10^2$ & 1   & off &off &  0.1 \\
          EVP1e2-SFRT1-VHR-SS$^\star$ & $2\times 10^7\quad$ ($10^5$) & $5 \,\,\,\qquad$ ($2\times 10^4$) & $10^2$ & 1   & off &off &  0.1 \\
          \hline
          EVP1e2-SFRT1-RT 	& $2\times 10^5\quad$ ($10^5$) & $500\quad$ ($2\times 10^4$) & $10^2$ & 1   & off &on &  1.0 \\
          EVP1e2-SFRT300-RT 	& $2\times 10^5\quad$ ($10^5$) & $500\quad$ ($2\times 10^4$) & $10^2$ & 300   & off &on &  1.0 \\
          \hline
          EVP1e2-SFRT6 	& $2\times 10^5\quad$ ($10^5$) & $500\quad$ ($2\times 10^4$) & $10^2$ & 6   & off &off &  1.0 \\
          EVP1e2-SFRT10 	& $2\times 10^5\quad$ ($10^5$) & $500\quad$ ($2\times 10^4$) & $10^2$ & 10   & off &off &  1.0 \\
          EVP1e2-SFRT20 	& $2\times 10^5\quad$ ($10^5$) & $500\quad$ ($2\times 10^4$) & $10^2$ & 20   & off &off &  1.0 \\
          EVP1e2-SFRT50 	& $2\times 10^5\quad$ ($10^5$) & $500\quad$ ($2\times 10^4$) & $10^2$ & 50   & off &off &  1.0 \\
          EVP1e2-SFRT100 	& $2\times 10^5\quad$ ($10^5$) & $500\quad$ ($2\times 10^4$) & $10^2$ & 100   & off &off &  1.0 \\
          \hline
          EVP1e2-SFRT1-SE$^{\ddag}$ & $2\times 10^5\quad$ ($10^5$) & $500\quad$ ($2\times 10^4$) & $10^2$ & 1   & on &off &  1.0 \\
          EVP1e2-SFRT300-SE$^{\ddag}$ 	& $2\times 10^5\quad$ ($10^5$) & $500\quad$ ($2\times 10^4$) & $10^2$ & 300   & on &off &  1.0 \\
          \hline
          \label{tab:runs}
      \end{tabular}}
      \begin{flushleft}
        \vspace{-0.5cm}
               {\small
                 $\phantom{}^{*}$ The LowT label indicates runs with low-temperature metal cooling for $Z=Z_\odot$.\\
                 $\phantom{}^{\diamondsuit}$ The IW label indicates runs with isotropic winds. The others are performed with axial winds.\\
                 $^\dag$ The LR label stands for low resolution.\\
                 $^\star$ The HR label stands for high resolution; the HR-SS label stands for high resolution and small softening; the VHR-SS label stands for very high resolution and small softening (see text in Sect.~\ref{sect:runs} for more details).\\
                 $\phantom{}^{\ddag}$ The SE label indicates runs with consistent stellar evolution calculations, and metal-dependent cooling in the temperature range $\sim 10-10^9\,\rm K$, for $Z$ evolving according to the proper stellar lifetimes and yields.
               }
      \end{flushleft}
\end{table*}



\begin{figure*}
  \begin{center}
    \includegraphics[width=0.52\textwidth]{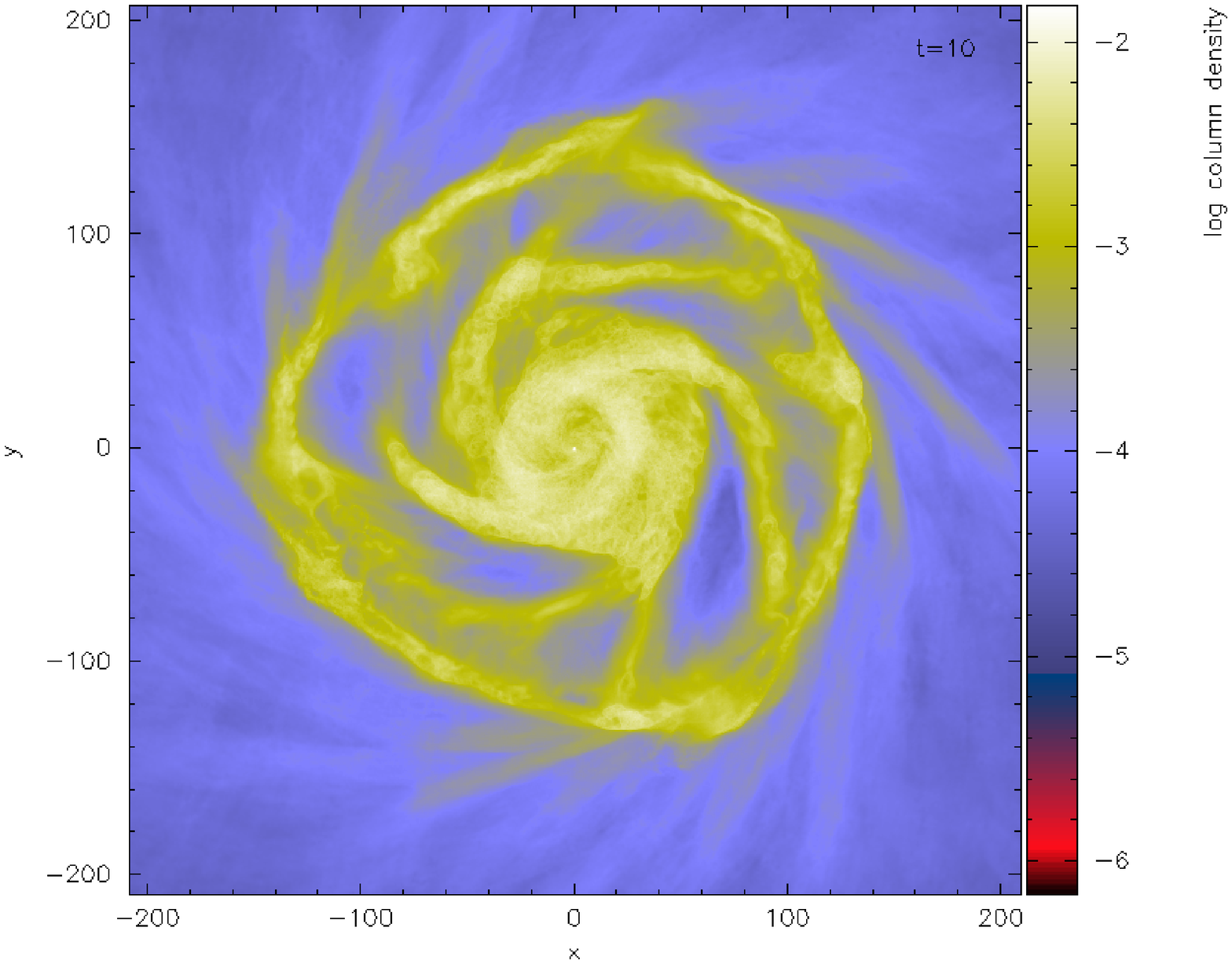}
    \hspace{-0.9cm}
    \includegraphics[width=0.52\textwidth]{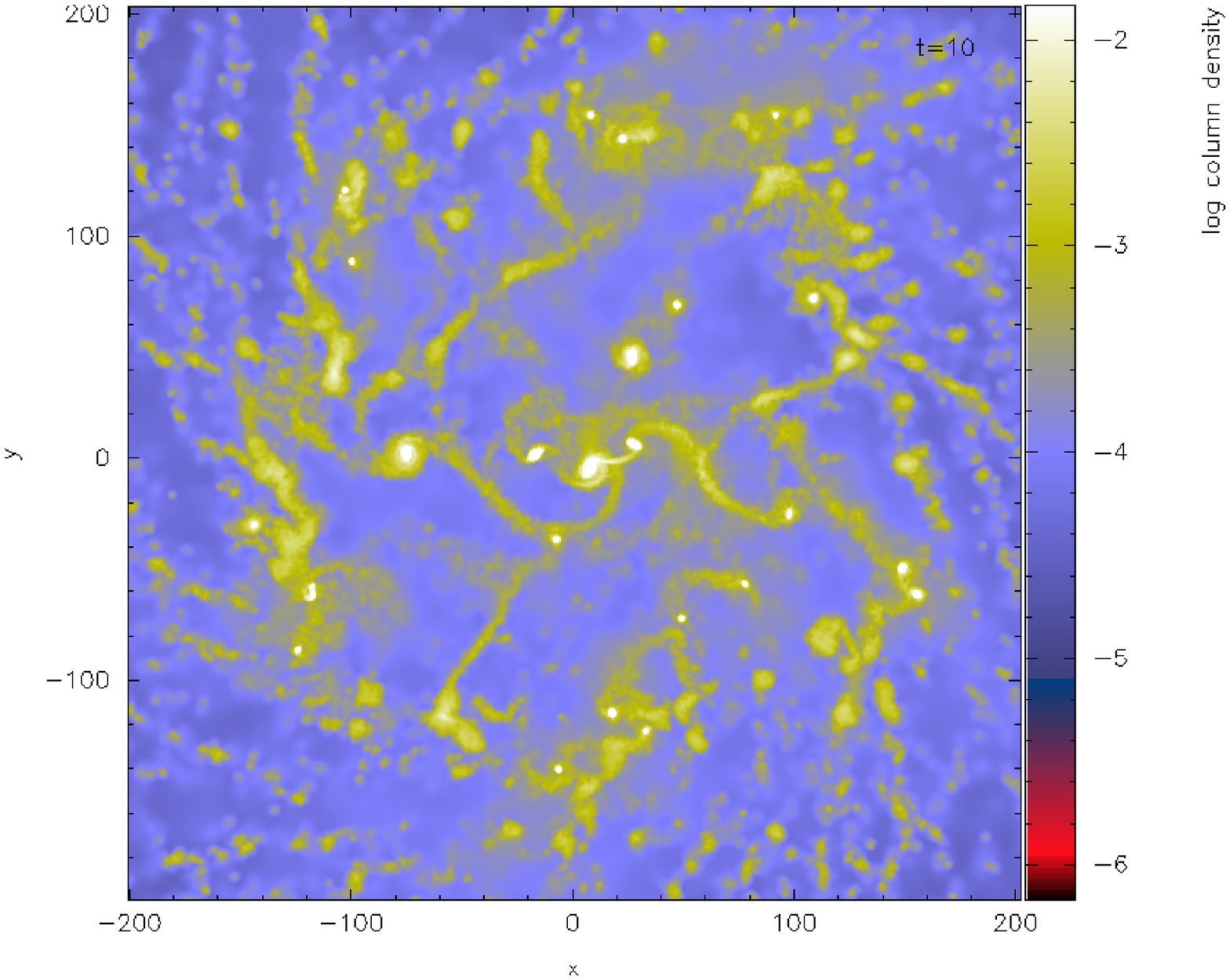}
  \end{center}
  \caption{\small
    Face--on view of the inner part of the CND at $t=10\,\rm Myr$.
    The color (logarithmic) scale refers to the gas surface density in units of $\rm 10^6\,M_\odot/pc^2$. Axes are in pc.
    {\it Left panel}: EVP1e2-SFRT1-VHR-SS.
    {\it Right panel}: EVP1e6-SFRT300-LowT.
  }
  \label{fig:maps}
\end{figure*}

\noindent
The runs we performed involve the study of several parameters and physical processes and we refer to Table~\ref{tab:runs} for the general outline of the features in the different simulations\footnote{
In the runs, we used 32 SPH neighbours for the density calculations, however, we checked the bahaviours up to 64 neighbours, as well, and found no significant difference in the results (mass profiles, star formation rates, angular momentum, relative angles, accretion rates, etc.).
}.
\\
The first, second, and third blocks refer to the simulations with axial wind feedback and evaporation factor A$_0=10^2$, A$_0=10^4$, and A$_0=10^6$, respectively.
For each case we run four simulations:
two with $\tsfr=1\,\rm Myr$ (one with metallicity $Z=0$, and H and He cooling down to $\sim 10^4\,\rm K$, and one with additional metal low-T cooling down to $\sim 10\,\rm K$ for an assumed constant $Z=Z_\odot$) and two with $\tsfr=300\,\rm Myr$ (one with $Z=0$, H and He cooling down to $\sim 10^4\,\rm K$, and one with additional metal low-T cooling down to $\sim 10\,\rm K$ with $Z=Z_\odot$).
\\
The corresponding complete, more detailed runs including (i) self-consistent metal production from stellar evolution according to the proper yields and lifetimes, (ii) exact abundance calculations, and (iii) metal-dependent cooling in the range $\sim 10-10^9\,\rm K$, will be also discussed in the following sections for the two limiting cases $\tsfr=1\,\rm Myr$ and $\tsfr=300\,\rm Myr$ (both with $A_0 = 10^2$).
\\
The resulting consequences from the cooling down to low temperatures on disk stability and fragmentation are readily seen in the density maps of figure~\ref{fig:maps}.
\\
In order to check the effects of outflows, we also run additional simulations with isotropic winds (IW), for the case with $A_0=10^2$, cooling down to $\sim 10^4\,\rm K$, and either $\tsfr=1\,\rm Myr$ or $\tsfr=300\,\rm Myr$.
\\
We carry out convergence studies over the dependence on  mass and spatial resolution by performing four additional runs, with $A_0=10^2$, $\tsfr=1\,\rm Myr$ and no low-T cooling.
The ``standard'' runs have a gravitational softening length of $1$ pc and $2 \times 10^5$ gas particles, with corresponding gas mass of 500 $\msun$.
By comparison, the low-resolution (LR) run has the same gravitational softening length, but ten times less gas particles. The two high-resolution (HR) simulations have 2 million gas particles (50 $\msun$ per particle) and are run with either a gravitational softening of $1$ pc or with a smaller softening of $0.1$ pc. Similarly, the very-high-resolution run (VHR) is run with a gravitational softening of 0.1 pc, but sampling our circum-nuclear disk with 20 million gas particles, corresponding to a gas mass resolution of 5 $\msun$.
\\
To check more realistic values of star formation timescales, we re-run the previous standard cases also with $\tsfr=6, 10, 20, 50, 100~\rm Myr$, besides the two $\tsfr=1~\rm Myr$ and $\tsfr=300~\rm Myr$ reference values.
\\
We then perform two simulations considering radiative transfer effects, as described in section~\ref{sect:implementation}, with standard resolution, assuming $A_0=10^2$ and either $\tsfr=1\,\rm Myr$ or $\tsfr=300\,\rm Myr$.
\\
Finally, we study the role of stellar evolution (SE) with consequent metal spreading by starting the simulations with a pristine gaseous environment ($Z=0$), by polluting it with metal yields \cite[][]{WW1995, vandenHoekGroenewegen1997, Thielemann2003}, from stellar evolution, according to the stellar lifetimes by \cite{PadovaniMatteucci1993}, and by consistently following the full chemical composition of the medium for H, He, O, C, Mg, S, Si, Fe.
For the technical details on this topic and studies on parameter dependencies, we refer the reader to \cite{Tornatore2004, Tornatore2007, Maio2007, Maio2010, Maio2011}.


\subsection{Star formation} \label{sect:SF}

\begin{figure*}
  \begin{center}
    \includegraphics[width=0.45\textwidth]{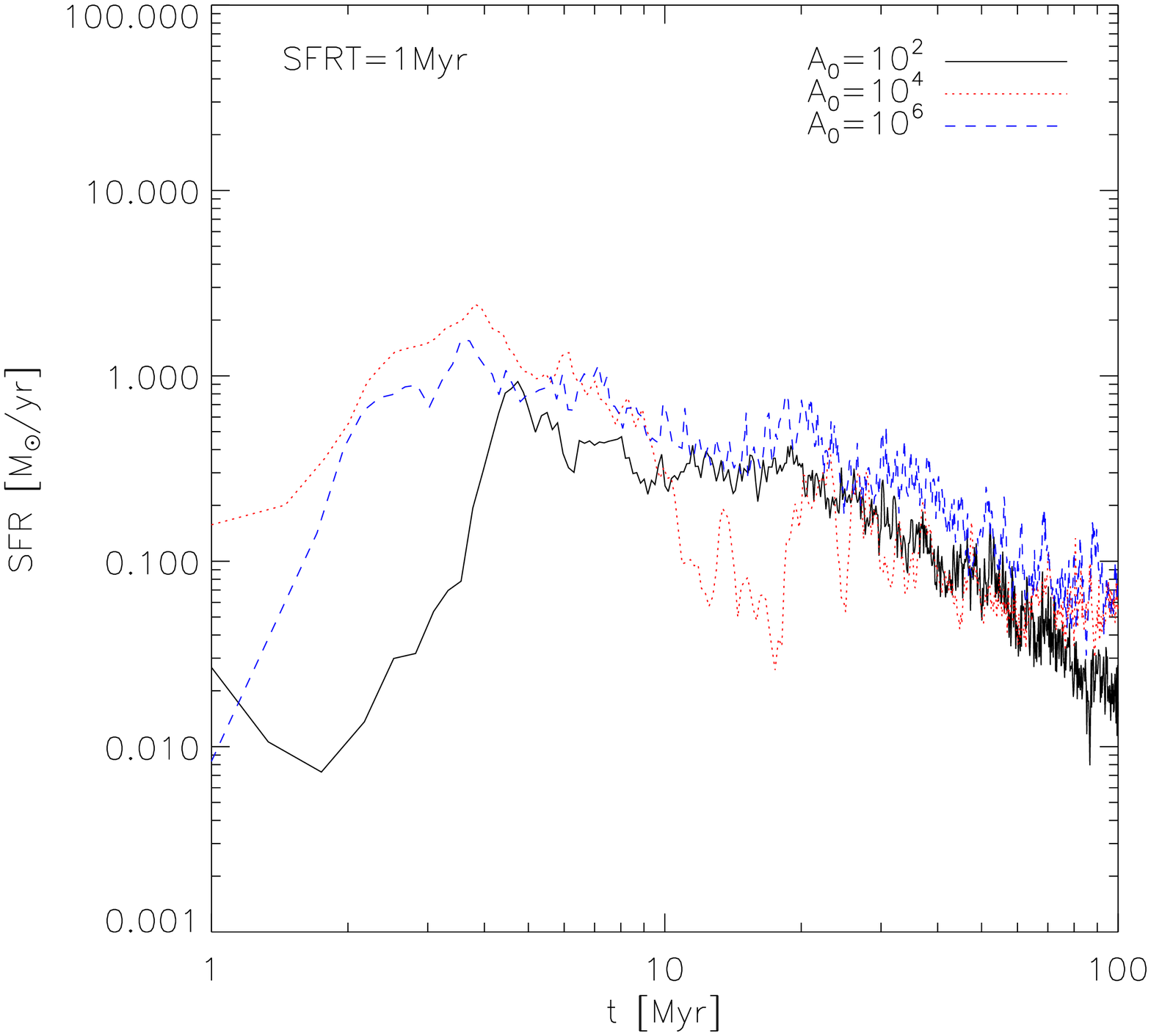}
    \includegraphics[width=0.45\textwidth]{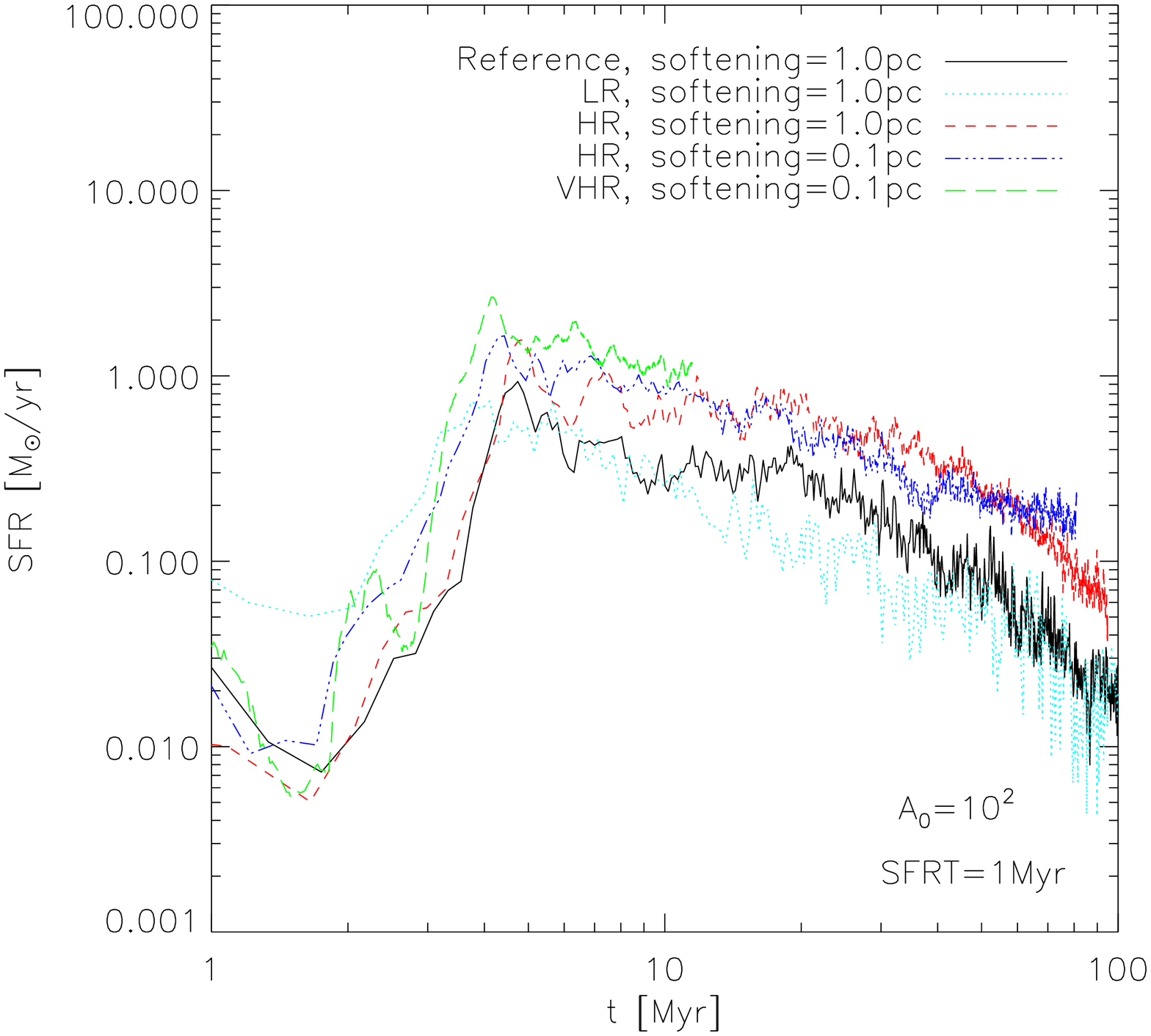}
  \end{center}
  \caption{\small
    Star formation rates for simulations evaporation parameters, $A_0$ (left), and resolution (right), as explained in the legends. The cases refer to $\tsfr=1\,\rm Myr$ with cooling down to $\sim 10^4\,\rm K$. The initial drop is due to the denser central regions of the CND that undergo star formation and get evacuated during the first few Myrs.
  }
  \label{fig:sfr}
\end{figure*}

\begin{figure*}
  \begin{center}
    \includegraphics[width=0.45\textwidth]{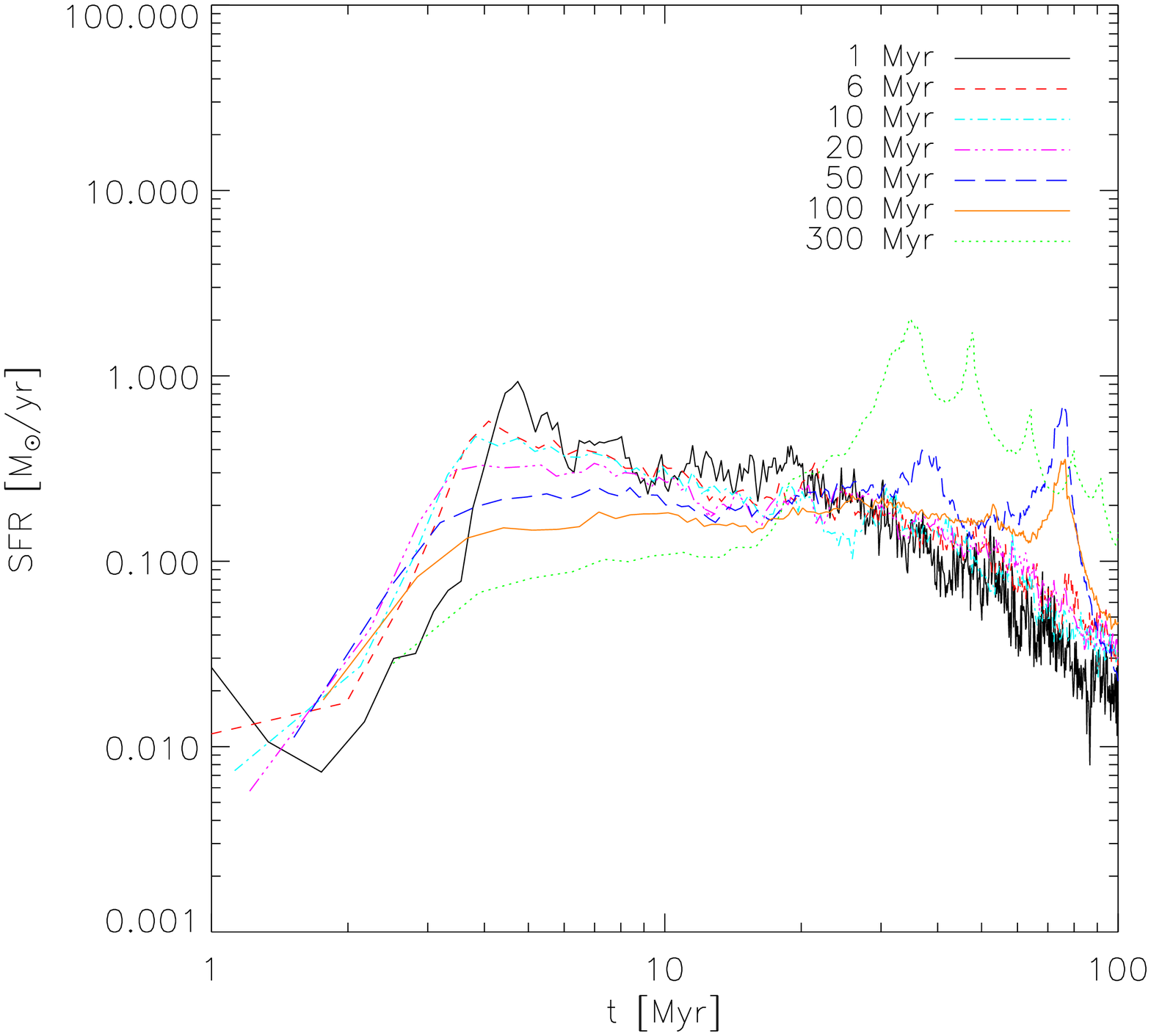}
    \includegraphics[width=0.45\textwidth]{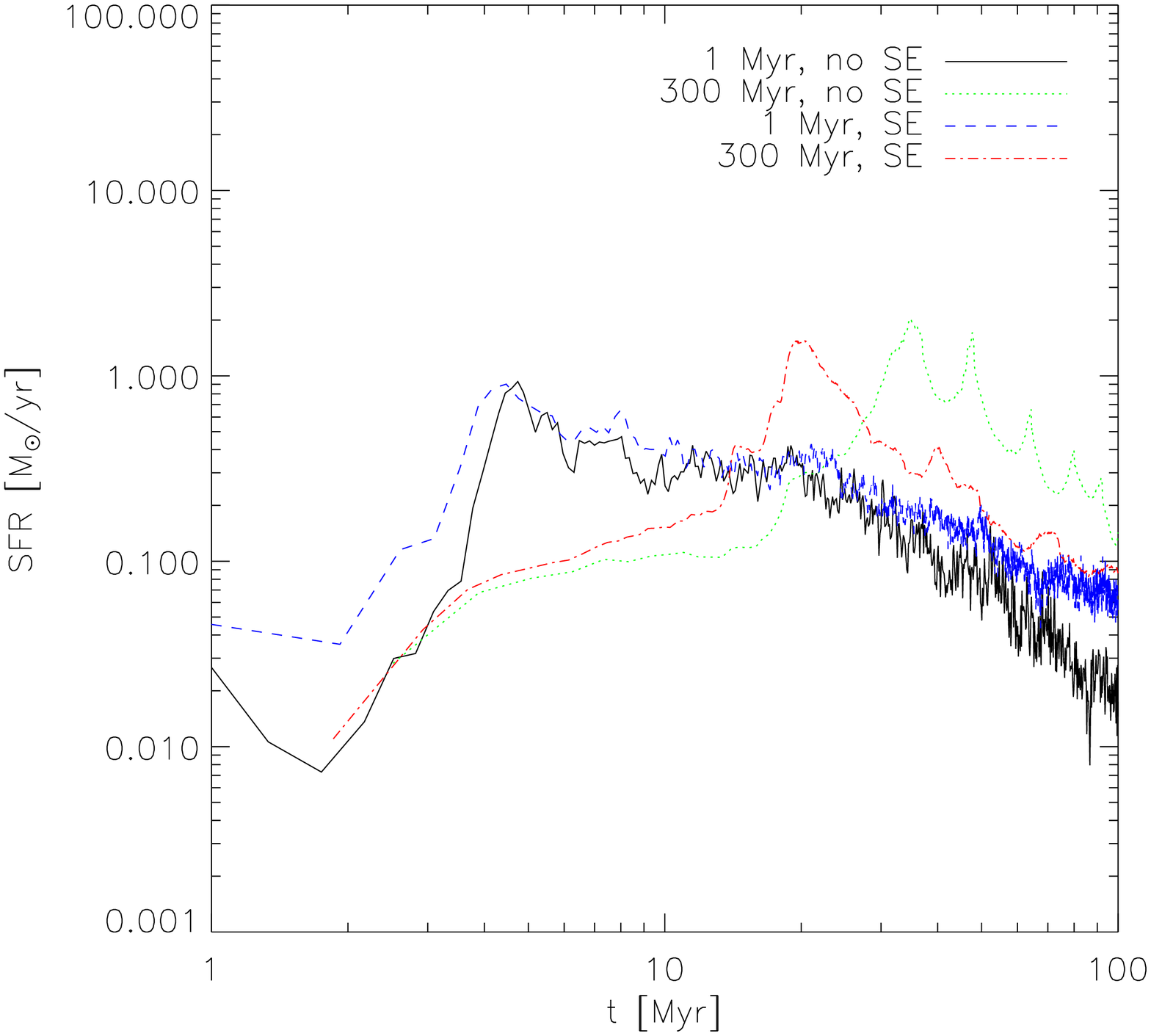}
  \end{center}
  \caption{\small
    {\it Left panel}:
    Star formation rates for simulations with different star formation time scales, $\tsfr = $~1, 6, 10, 20, 50, 100, 300~Myr.
    {\it Right panel}: Star formation rates for simulations with inclusion of stellar evolution (SE) treatment and metal pollution process compared to the corresponding star formation rates in the runs with no SE treatment, for $\tsfr=1\,\rm Myr$ and $\tsfr=300\,\rm Myr$, with $A_0 = 10^2$.
  }
  \label{fig:sfr2}
\end{figure*}

\begin{figure*}
  \begin{center}
    \includegraphics[width=0.45\textwidth]{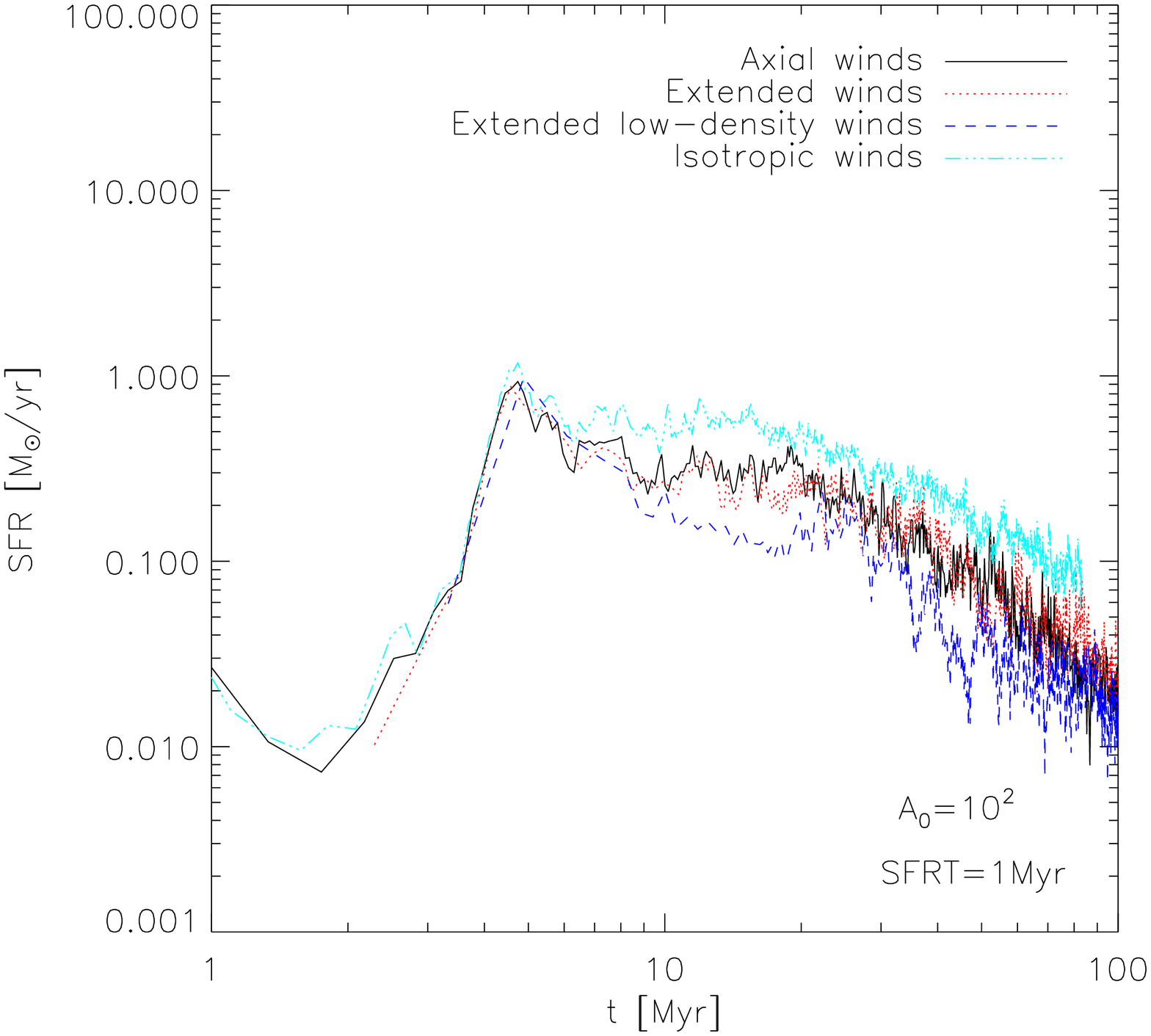}
    \includegraphics[width=0.45\textwidth]{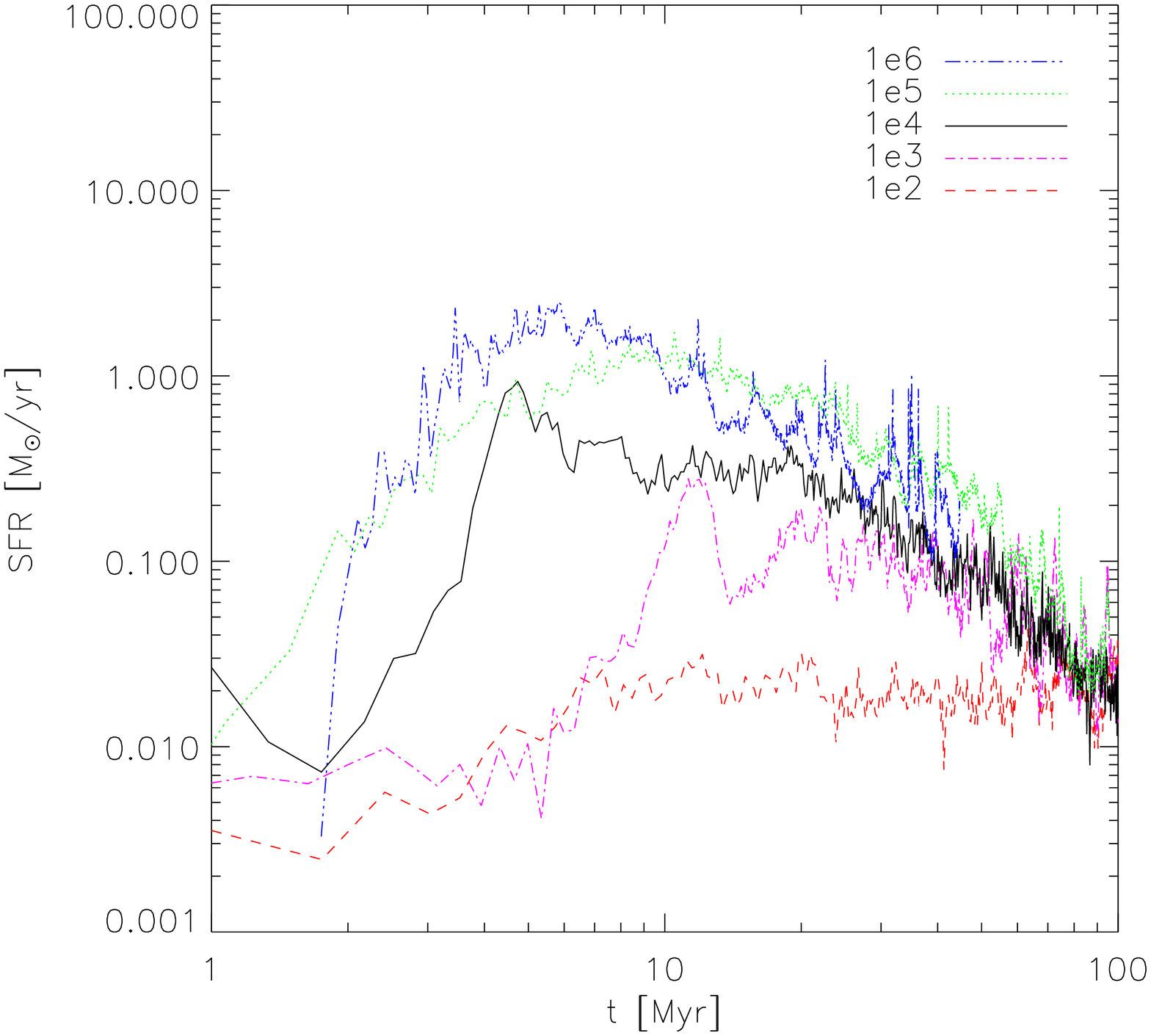}
  \end{center}
  \caption{\small
    {\it Left panel}:
    Star formation rates for simulations with different wind parameters for the $\tsfr=1\,\rm Myr$ and $A_0 = 10^2$ case: 
    axial winds with coupling with the ISM after a travel length of 1~pc (black solid line, simply labeled as "Axial winds");
    axial winds with coupling with the ISM after a travel length of 10~pc (red dotted line, labeled as "Extended winds");
    axial winds with coupling with the ISM after a travel length of 10~pc and density contrast below 0.1 (blue dashed line, labeled as "Extended low-density winds");
    isotropic winds (cyan dot-dashed line).
    {\it Right panel}:
    Star formation rates for simulations with gas density thresholds of
    $10^2 - 10^6 \,\rm cm^{-3}$, as indicated by the labels, for the $\tsfr=1\,\rm Myr$ and $A_0 = 10^2$ case with axial winds.
  }
  \label{fig:sfr3}
\end{figure*}

\noindent
In all our runs, independently of the resolution, the softening, or other parameters, the disk experiences a burst of star formation, with a peak of a few solar masses per year, lasting for $\sim 1 - 10\,\rm Myr$ (see figure~\ref{fig:sfr}).
These values are compatible, although on the low side, with the typical values of star formation rates in circum-nuclear disks associated with star-bursts and late stages of galaxy interactions, from a few solar masses per year up to several hundreds \cite[][]{Kennicutt1998, Downes1998, Davies2003, Davies2004, Evans2006, DownesEckart2007}.
After $t\approx 100\,\rm Myr$, the star formation rate decreases by 1--2 orders of magnitude, depending on the particular parameters used, but it always peaks at $3-5\,\rm Myr$ for $\tsfr=1\,\rm Myr$, and the global behaviour is quite independent of $A_0$ (left panel of figure~\ref{fig:sfr}) and of resolution (right panel of figure~\ref{fig:sfr}).
The cases with $\tsfr=300\,\rm Myr$ result in a slightly later peak of star formation at $\sim 5-30\,\rm Myr$
with slightly larger values.
This is due to the longer timescale considered and the resulting higher clumpiness reached in the star forming regions.
However, in all the simulations the central region of the disk is efficiently converted into stars, that become a relevant component after $\sim 10\,\rm Myr$.
\\
In figure~\ref{fig:sfr2}, we also show the star formation rates for different $\tsfr$ assumptions: $\tsfr=$~1, 6, 10, 20, 50, 100, 300~Myr (left panel)\footnote{
We note that a value of $\tsfr=1\rm Myr$ (with gas threshold of $10^4\,\rm cm^{-3}$) would correspond to star formation efficiencies of roughly $\sim 50\%$, while $\tsfr=300\rm Myr$ would correspond to a very low value of $\sim 0.15\%$.
As mentioned in the Introduction, these are quite extreme numbers, compared to expected values of a few per cents, or so.
A more realistic assumption of star formation efficiency of $\sim 1\%$ would require $\tsfr\sim 50\,\rm Myr$.
Even though there are some cases (that we also take into account) for which timescales can be significantly shorter, as in the Milky Way, for example, for which it is expected $\sim 5-6\,\rm Myr$ \cite[][]{Burkert2006}, or the T-Tauri complex for which observed timescales are of the order of $\sim 1-3\,\rm Myr$ \cite[][]{PallaStahler2000, Hartmann2001}.
},
and additional calculations including stellar evolution and metal production (right panel).
We see (left panel of figure~\ref{fig:sfr2}) that when increasing $\tsfr$ -- i.e. when lowering star formation efficiency -- the SFRs during the initial phases (i.e., by $\lesssim 10\,\rm Myr$) get shallower and the bulk of formation of stars gets gradually shifted to later times.
After the first burst, efficiently star forming models with short $\tsfr$ (say around $\lesssim 50 \,\rm Myr$) show a clearly declining trend, due to the lack of remaining cold gas, that has been rapidly consumed during the initial phases. Feedback processes take place immediately, heat the surrounding medium, and inhibit further star formation.
In less efficiently star forming models, with larger $\tsfr$ values, we assist to a slowly increasing trend of the SFRs for the initial tens Myrs. In these scenarios, gas can cool and condense for longer time without forming significant amounts of stars.
At $\gtrsim 20\,\rm Myr$, cooling processes finally lead to a number of strong SFR peaks, that take place when the medium is, in average, colder and more clumpy than in the previous cases.
Thus, only after these episodes feedback effects become efficient in heating the gas and halting further stellar production.
Interestingly, a limiting situation is represented by $\tsfr = 50\,\rm Myr$, for which the SFR is roughly flat over the entire lapse of time simulated, and reflects an almost continuous star formation process during which feedback effects and gas cooling are always well regulated.\\
The inclusion of detailed stellar evolution with self-consistent metal production (right panel of figure~\ref{fig:sfr2}) determines some difference in the time evolution of the SFR due to the increasing metallicities that slightly boost star formation. However, no major effects in the final results are found.
\\
In figure~\ref{fig:sfr3}, we finally plot the evolution of the SFRs in models with different wind shapes (left) and different gas density thresholds (right).
On the left panel, we show results from simulations with $\tsfr=1\,\rm Myr$ and $A_0 = 10^2$, considering:
axial winds with coupling with the ISM after a travel length of 1~pc (black solid line, simply labeled as "Axial winds");
axial winds with coupling with the ISM after a travel length of 10~pc (red dotted line, labeled as "Extended winds");
axial winds with coupling with the ISM after a travel length of 10~pc and density contrast below 0.1 (blue dashed line, labeled as "Extended low-density winds");
isotropic winds (cyan dot-dashed line).
While the first and second case are basically very similar, some slight difference is found for the third case, where axial winds start interacting with the medium only in under-dense regions, and hence, lead to less star formation after the first few Myrs.
In the case of isotropic winds, particles are not collimated in one direction, but can interact with the surrounding gas at $4\pi$, compressing the gas in all directions, hence, a larger SFR is obtained.
Quantitatively, the boost to the SFR due to isotropy is only a factor of a few, though.
On the right panel, SFRs for the same initial setup and axial winds, but with gas density thresholds of $10^2-10^6\,\rm cm^{-3}$ are plotted.
Since star formation is strongly dependent on density, larger thresholds determine higher SFRs and lower thresholds determine lower SFRs.
In the cases considered, differences can reach up to $\sim 2$ orders of magnitude, ranging from $\lesssim 10^{-2}\,\rm M_\odot/yr$ (for a threshold of $10^2\,\rm cm^{-3}$) up to several $ \rm M_\odot/yr $ (for a threshold of $10^6\,\rm cm^{-3}$).
The obvious consequence is that the little amount of stars formed in low-density-threshold scenarios will trigger feedback processes acting essentially on low-density gas, and that could be also effective in randomizing the dynamics of the medium (see more discussion in the parameter study later).

\section{Spin}\label{sect:spin}

\noindent
We take advantage of the extremely wide variety of our simulations to track the dynamics of the stream of particles inflowing towards the center and show the results for different cases.

\subsection{Method}
\noindent
We estimate the accretion rate and angular momentum of the inflowing material through the smallest surface surrounding the SMBH that can be numerically resolved\footnote{
We note that the sphere of influence (extending out to $\sim 20\,\rm pc$) is largely resolved in all our runs.
}
(e.g., 3~pc and 0.3~pc for simulations with a 1~pc and 0.1~pc resolution, respectively).
Due to the large range of scales involved, the formation of accretion disks around the SMBH could not be resolved down to $\lsim 0.01$ pc, where the gas can be modelled as a standard accretion disk.
Therefore, the relevant properties of the disk and their evolution in time have been simply inferred from the properties of the particles inflowing through the previously mentioned surface.
In particular, we assume that: 1) the accretion rate onto the hole is equal to the accretion rate at the smallest resolved scales; 2) the direction of the orbital angular momentum in the external part of the unresolved accretion disk is parallel to the direction of the total angular momentum (relative to the SMBH) of the infalling gas at the resolution limit.
We notice that all these assumptions are necessary working hypotheses.
Nevertheless, we stress that these information can be gathered only by performing high-resolution numerical simulations that accurately resolve the ``sphere of influence'' of the black hole.
We limit the gas inflow to the Eddington rate for the SMBHs, and use the SMBH accretion histories obtained from our SPH simulations to evolve the SMBH spin vector.
We notice that while the SMBH spin and mass evolve in the semianalytical model we use to postprocess our simulations (described in section~\ref{sect:spin method}), the SMBH mass is constant in our SPH simulations, and the gas flowing toward the SMBH is not removed from the simulations.
As shown in section~\ref{sect:spin results}, we limit our simulations to $t\lsim 20~\rm Myr$, to limit the mis-match between the SMBH mass in the simulations and in the semianalytic model to a factor of a few.
We stress that such a small mass growth would not change substantially the gas dynamics, since the CND is significantly more massive than the SMBH in our initial set-up.
\\
If the orbital angular momentum of the disk around the SMBH is misaligned with respect to the SMBH spin, the coupled action of viscosity and relativistic Lense-Thirring precession warps the disk in its innermost region \cite[][]{bardeenpetterson1975} over a timescale shorter than the viscous/accretion timescale \cite[][]{Scheuer1996, NatarajanPringle1998, Bogdanovic2007}.
We develop an analytical framework to model the response of the accretion disk (in the innermost non-resolved region) to the angular momentum of the mass inflow as described below.
At each timestep we can calculate the evolution of the magnitude and direction of the SMBH spin \cite[][]{Perego2009, Dotti2010}.


\subsection{Bardeen-Petterson effect}\label{sect:spin method}
\noindent
We use the SMBH accretion histories obtained from our SPH simulations to follow the evolution of each SMBH spin vector, ${\mathbf J}_{\rm BH}=(aGM_{\rm BH}^2/c){{\mathbf j}}_{\rm BH}$, where $0 \leq a \lesssim 1$ is the dimensionless spin parameter and ${\mathbf j}_{\rm BH}$ is the spin unit vector. We will use the term maximally rotating for a black hole with $a=0.998$, following \cite{Thorne1974} who showed that accretion driven spin-up is limited to such value.
Magnetic fields connecting material in the disk and the plunging region may further reduce the equilibrium spin. Magneto-hydrodynamic simulations for a series of thick accretion disks suggest an asymptotic equilibrium spin at $a\approx 0.9$ \cite[][]{Gammie2004}.
\\
The scheme we adopt to study spin evolution is based on the model developed in \cite{Perego2009}.
Here we summarize the algorithm.
We assume that inflowing gas forms a geometrically thin/optically thick $\alpha$-disk \cite[][]{ShakuraSunyaev1973, ShakuraSunyaev1976} on sub-parsec scales (not resolved in the simulation), and that the outer disk orientation is defined by the unit vector ${\mathbf l}_{\rm edge}$.
The evolution of the $\alpha$-disk is related to the radial viscosity, $\nu_1$, and the vertical viscosity, $\nu_2$: 
$\nu_1$ is the radial shear viscosity, while $\nu_2$ is the vertical shear viscosity associated to the diffusion of vertical warps through the disk.
The two viscosities can be described in terms of two different dimensionless viscosity parameters, $\alpha_1$ and $\alpha_2$, through the relations
\begin{equation}
  \nu_{1,2} = \alpha_{1,2}Hc_{\rm s},
\end{equation}
where $H$ is the disk vertical scale height and $c_{\rm s}$ is the sound speed of the gas in the accretion disk.
Both of them depend on $R$, $M_{\rm BH}$, $\dot{M}$ and $\alpha_1$.
Here we assume the \cite{ShakuraSunyaev1973} solutions for the gas pressure dominated zone.
\\
We further assume \cite[][]{LodatoPringle2007} 
\begin{equation}
  \alpha_2=f_2 / (2 \alpha_1),
\end{equation}
with\footnote{
These values are slightly different from others also found in literature \cite[as e.g. in][]{BateLodatoPringle2010}.
}
\begin{equation}
  \alpha_1 = 0.1
\end{equation}
and $f_2 = 0.6$, i.e.
\begin{equation}
  \alpha_2 = 3,
\end{equation}
and power law profiles for the two viscosities
\cite[][]{ShakuraSunyaev1973,ShakuraSunyaev1976}, 
\begin{equation}
  \label{norm}
  \nu_{1,2} \propto R^{3/4}.
\end{equation}
\noindent
If the orbital angular momentum of the disk around the SMBH is misaligned with respect to the SMBH spin, the coupled action of viscosity and relativistic Lense-Thirring precession warps the disk in its innermost region forcing the fluid to rotate in the equatorial plane of the spinning SMBH \cite[][]{bardeenpetterson1975,Kumar1985}.
The timescale of propagation of the warp is short compared with the viscous/accretion timescale \cite[][]{Scheuer1996, NatarajanPringle1998}, so that the deformed disk reaches an equilibrium profile that can be computed by solving the equation of conservation of angular momentum
\begin{eqnarray}
  \label{eqn:angular momentum}
  \frac{1}{R}\frac{\partial}{\partial R}(R {\mathbf L} v_{\rm R}) =
  \frac{1}{R}\frac{\partial}{\partial R}\left(\nu_1 \Sigma R^3 \frac{d\Omega}{dR}~ {\mathbf  l} \right)+ \nonumber \\
  +\frac{1}{R}\frac{\partial}{\partial R}\left(\frac{1}{2}\nu_2 R L \frac{\partial {\mathbf  l}}{\partial R} \right)
  + \frac{2G}{c^2} \frac{{\mathbf J}_{\rm BH} \times {\mathbf L}} {R^3}
\end{eqnarray}
where $v_R$ is the radial drift velocity, $\Sigma$ is the surface density, $\Omega$ is the Keplerian velocity of the gas in the disk, and $\mathbf{L}$ is the local angular momentum surface density of the disk, defined by its modulus $L$ and the versor ${\mathbf l}$ that defines its direction.\\
The boundary conditions to eq.~(\ref{eqn:angular momentum}) are the direction of ${\mathbf L}$ at the outer edge ${\mathbf l}_{\rm edge}$, the mass accretion rate (that fixes the magnitude of $\Sigma$), and the values of mass and spin of the massive BH.
\\
The values of ${\mathbf l}_{\rm edge}$ and of the mass accretion rate are directly extracted from the simulations:
in particular, they are computed considering those SPH particles nearing the MBH gravitational sphere of influence that are accreted.
\\
The spin value, $\rm J_{BH}$, is treated on post-processing in a Monte Carlo fashion, by randomly assuming several different realizations at the initial time, and evolving each of them.
More exactly, the direction of the SMBH spin changes in response to its gravito-magnetic interaction with the disk on a timescale longer than the time scale of warp propagation \cite[][]{Perego2009}.
This interaction tends to reduce the degree of misalignment between the disk and the SMBH spin, decreasing with time the angle between ${\mathbf J}_{\rm BH}$ and ${\mathbf l}_{\rm edge}$ \cite[][]{Bogdanovic2007}.
So, the SMBH spin evolution is followed by solving for the equation
\begin{equation}
  \label{eqn:jbh precession-disk}
  \frac{d{\mathbf J}_{\rm BH}}{dt} =\dot{M}\Lambda(R_{\rm ISO}) {\mathbf l}(R_{\rm ISO}) + \frac{4\pi G}{c^2}\int_{\rm disk}\frac{{\mathbf L} \times {\mathbf J}_{\rm BH}}{R^2}dR.
\end{equation}
The first term in eq.~(\ref{eqn:jbh precession-disk}) accounts for the angular momentum deposited onto the SMBH by the accreted particles at the innermost stable orbit (ISO), where $\Lambda(R_{\rm ISO})$
denotes the specific angular momentum at $R_{\rm ISO}$ and $ {\mathbf l}(R_{\rm ISO})$ the unit vector parallel to ${\mathbf J}_{\rm BH}$,  describing the warped shape of the disk.
The second term instead accounts for the gravo-magnetic interaction of the SMBH spin with the warped disk. It modifies only the SMBH spin direction (and not its modulus), conserving the total angular momentum of the composite (MBH+disk) system \cite[][]{King2005}.
The integrand peaks at the warp radius ($R_{\rm warp}$), where the disk deformation is the largest\footnote{
The exact definition of $R_{\rm warp}$ is where the vertical viscous timescale $t_{\nu_2}\simeq R^2/\nu_2$ in the disk is comparable to the Lense-Thirring precession timescale. Because $R_{\rm warp}$ and the radius at which the disk is maximally deformed are  comparable \cite[][]{Perego2009}, we simplify the notation in the paper using only $R_{\rm warp}$.
}.
The equation incorporates two timescales: the accretion time related to the first right-hand term describing the $e-$folding increase of the spin modulus, and the shorter timescale of SMBH spin alignment
\begin{equation}
  \tau_{\rm al} \sim  10^5 a^{5/7} \left(\frac{M_{\rm BH}}{4 \times 10^6 \msun}\right)^{-2/35} f_{\rm Edd}^{-32/35} {\rm yr},
      \label{eq:alignts}
\end{equation}
that will ensure a high degree of SMBH-disk gravito-magnetic coupling during SMBH inspiral.
In Eq~(\ref{eq:alignts}), $f_{\rm Edd}$ is the SMBH luminosity in units of $L_{\rm Edd}$.
\\
We apply iteratively eq.~(\ref{eqn:angular momentum}) and (\ref{eqn:jbh precession-disk}), by using inputs from the SPH simulations that give the values of the mass accretion rate, the SMBH
mass and the direction of ${\mathbf l}_{\rm edge}$.  The algorithm returns, as output, the spin vector, that is, its magnitude and direction.
At each timestep our code therefore provides the angle between the spin vector of the SMBH and the global angular momentum vector of the circum-nuclear disk.


\subsection{Resulting spin alignment and magnitude} \label{sect:spin results}

\begin{figure*}
  \begin{center}
    \includegraphics[width=0.49\textwidth]{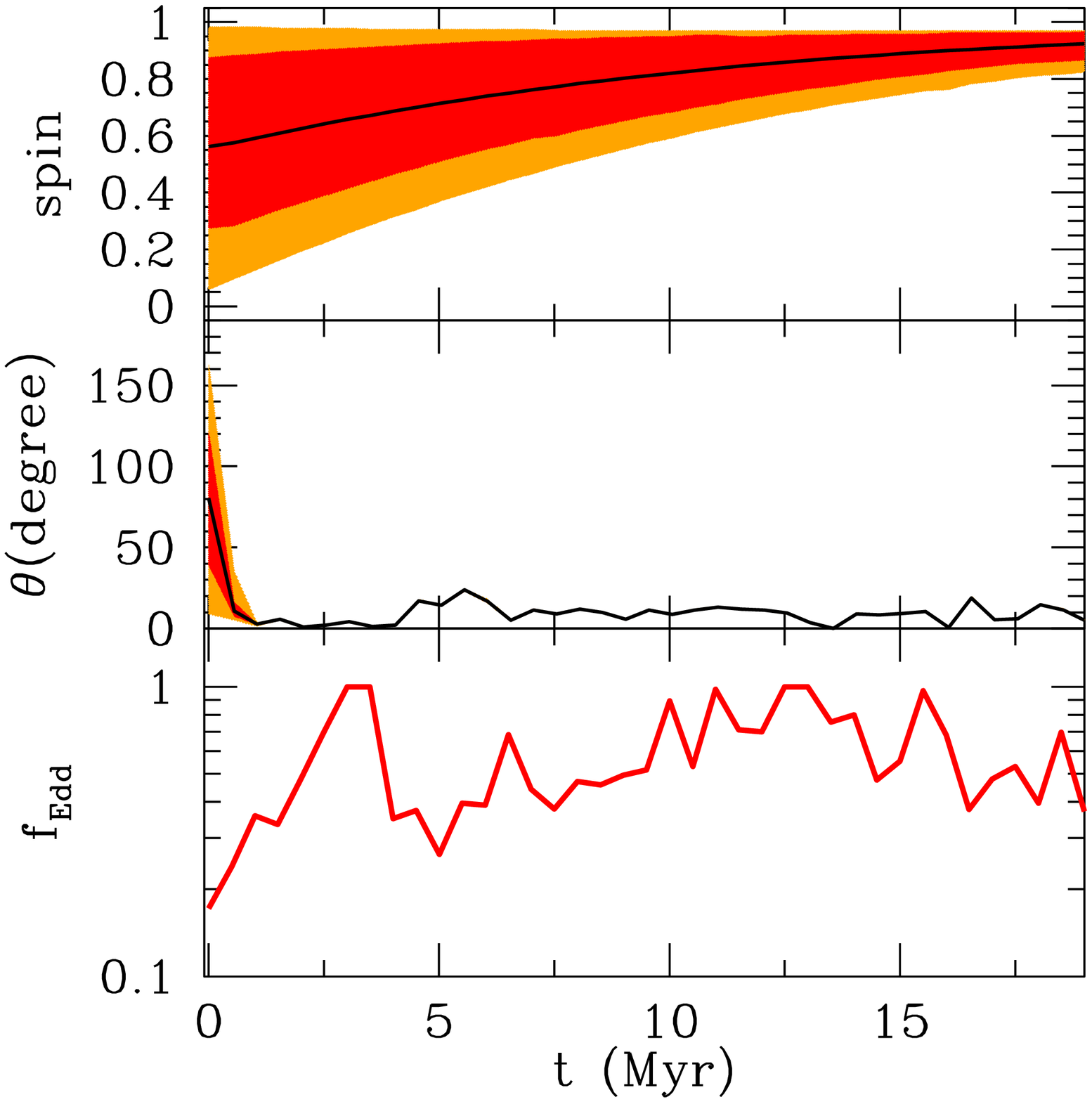}
    \includegraphics[width=0.49\textwidth]{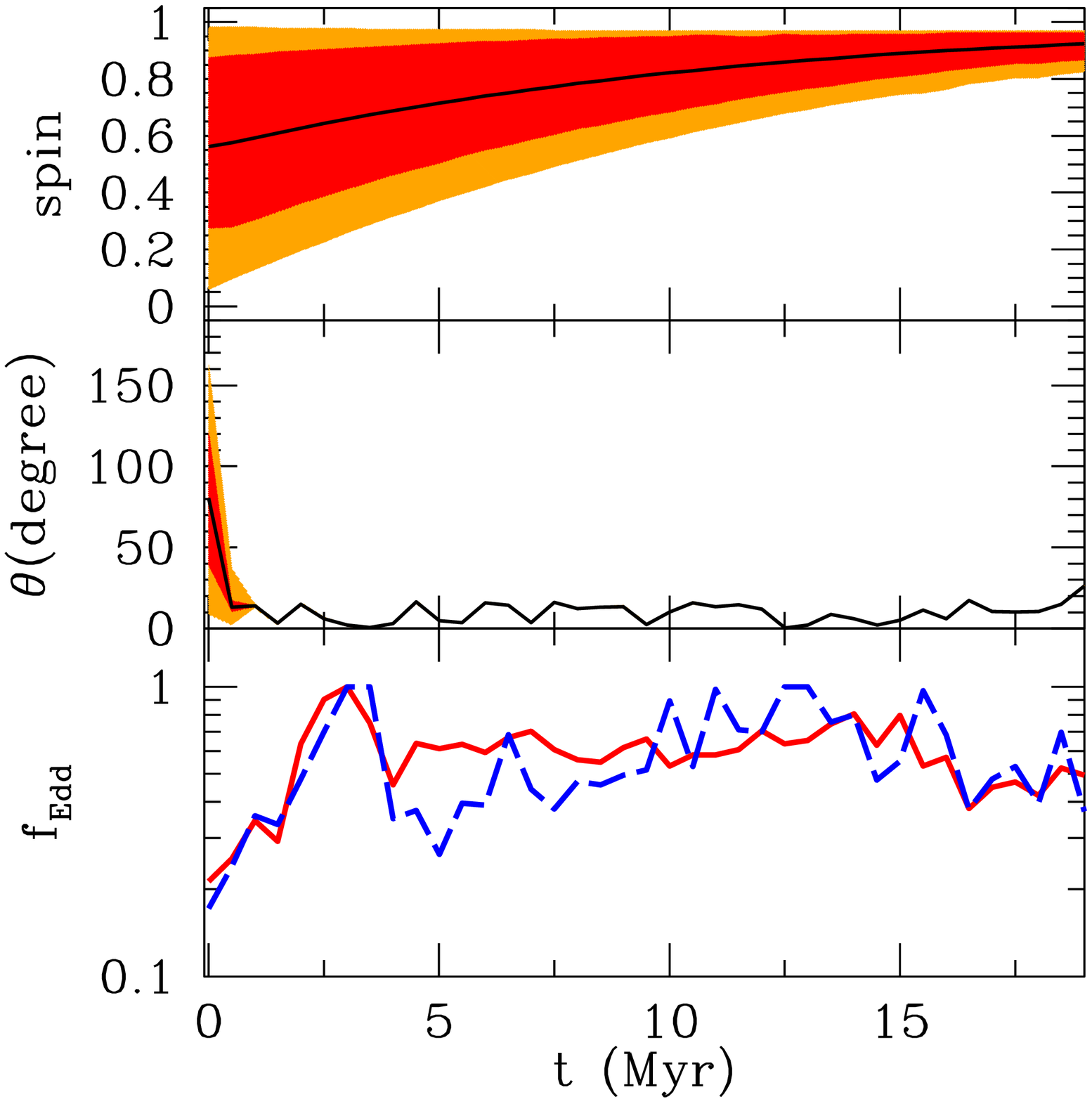}
    \includegraphics[width=0.49\textwidth]{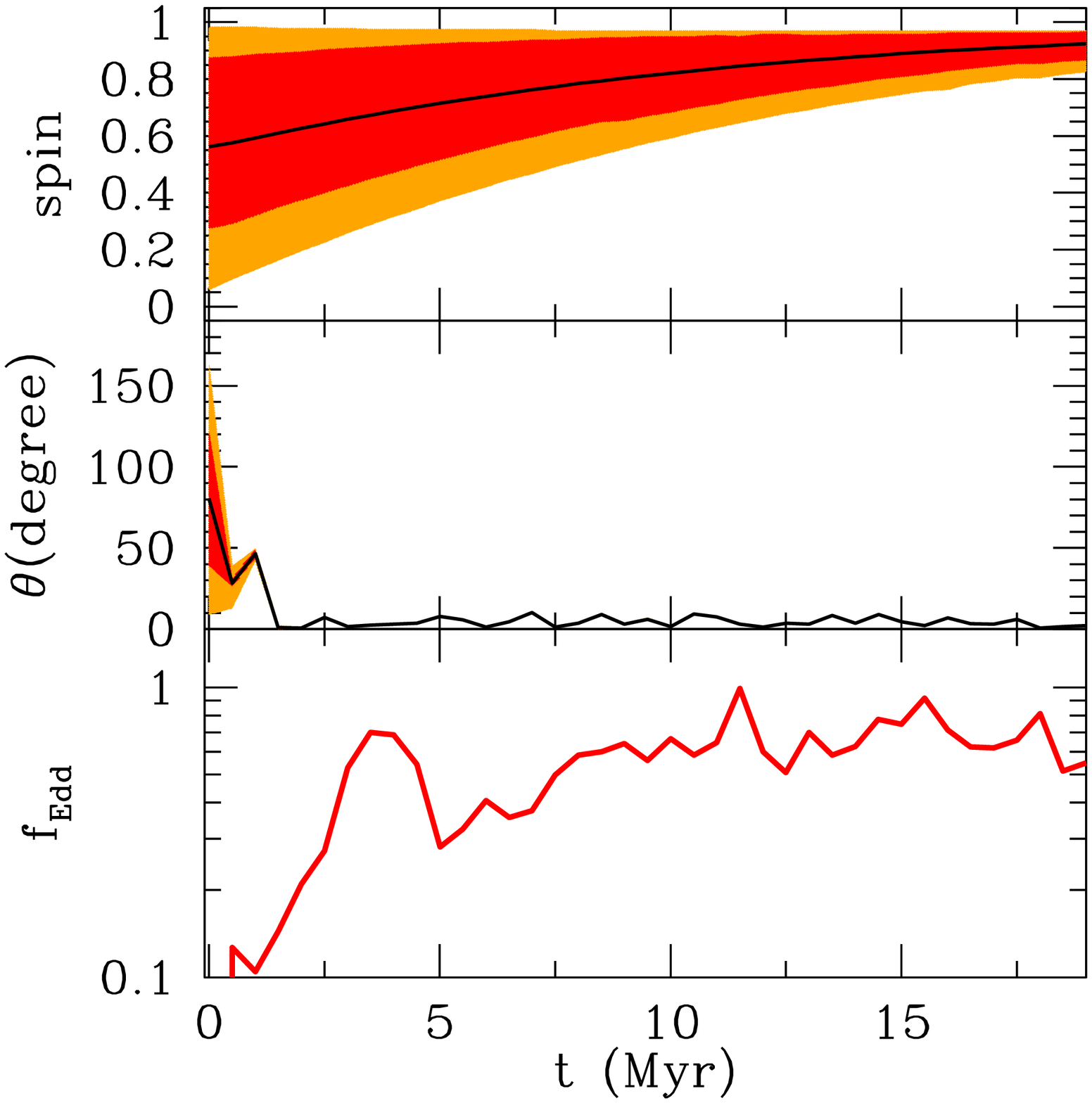}
    \includegraphics[width=0.49\textwidth]{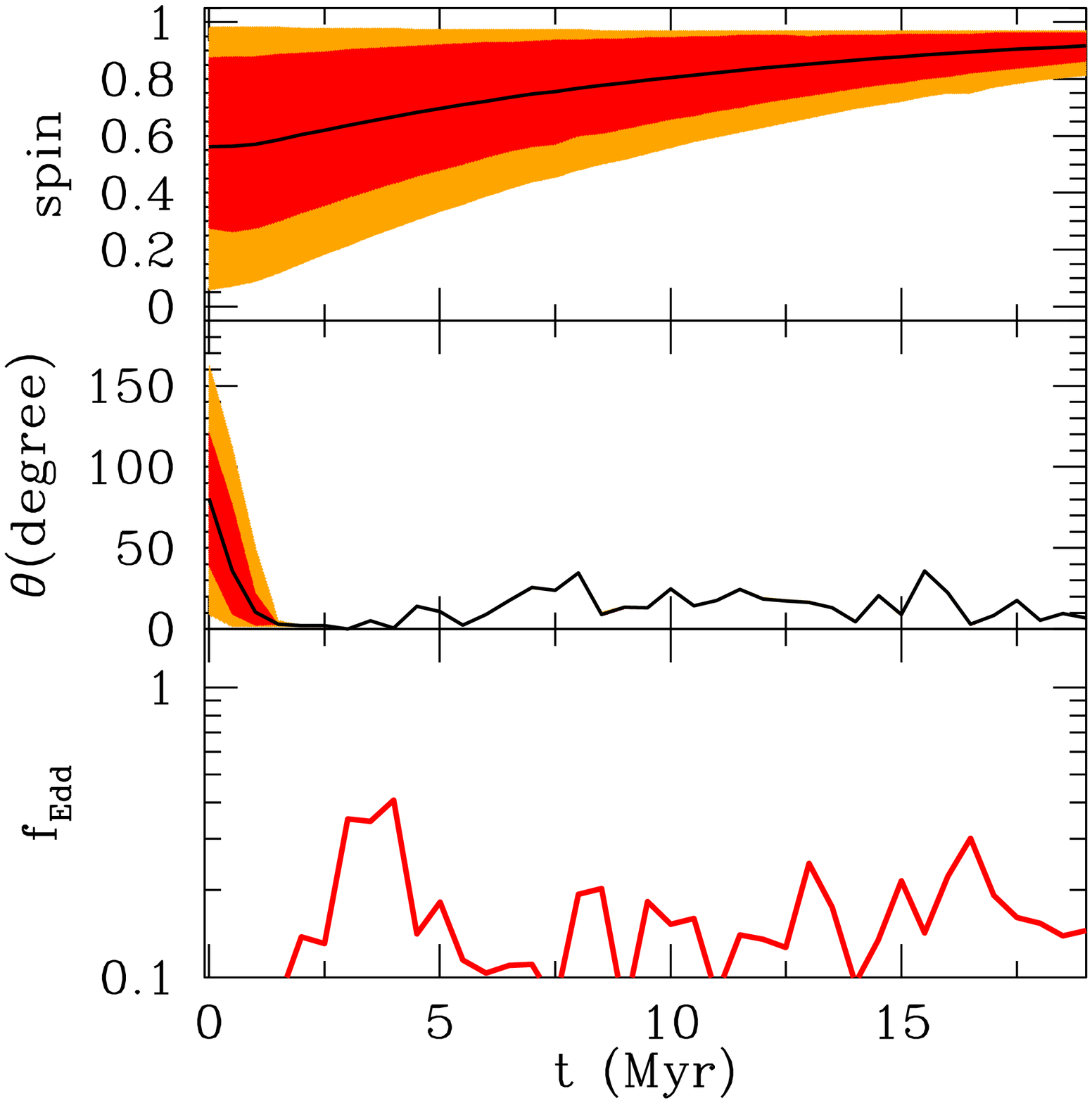}\\
  \end{center}
  \caption{\small
    For each panel, the time evolution of the spin parameter (top), of the spin orientation angle (middle), and of the luminosity in units of the Eddington luminosity, $f_{\rm Edd}$ (bottom), computed at a distance of 3~pc from the central BH, are shown.  In the upper and middle panels the solid lines refer to the mean values of the spin magnitude and orientation $\theta$, averaged over 100 realizations. When $\theta$ approaches zero, the accreted material is aligned with the SMBH spin (``coherent'' accretion). When $\theta$ is large, material is flowing with angular momentum misaligned with respect to the spin, causing the spin to decrease. The darker and lighter (red and orange in the online version) shaded areas refer to the regions encompassing 1- and 2-sigma around the means. The different panels refer to the runs with $500\,\rm M_\odot$ mass resolution: EVP1e2-SFRT1 (top left), EVP1e2-SFRT1-LowT (top right), EVP1e2-SFRT1-IW (bottom left), EVP1e2-SFRT1-RT (bottom right). In the top right panel the evolution of $f_{\rm Edd}$ for the EVP1e2-SFRT1 run is shown with a dashed line to facilitate the comparison.
  }
  \label{fig:spins}
\end{figure*}

\begin{figure}
  \centering
  \includegraphics[width=0.49\textwidth]{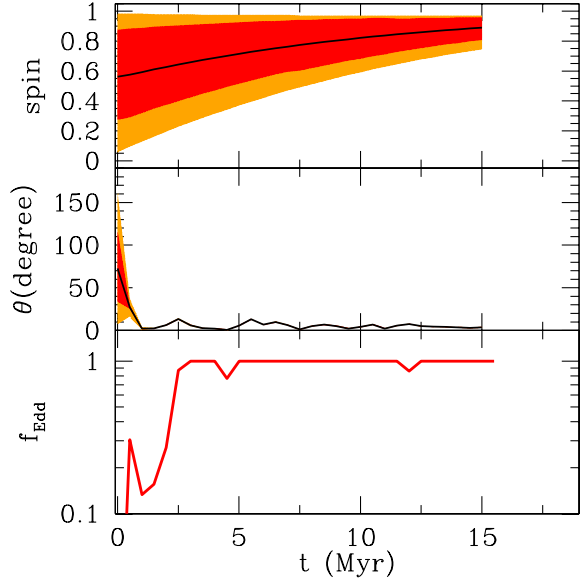}
  \caption{\small
    Same as figure~\ref{fig:spins} for the highest resolution run (simulation EVP1e2-SFRT1-VHR-SS).  
  }
  \label{fig:hires}
\end{figure}

\noindent
We will now discuss the results expected for the spin behaviour in the various runs presented before. We will start with general trends  and then show also one of the highest resolution runs.
\\
Figure~\ref{fig:spins} shows the time evolution of the spin magnitudes (top panel).
We show here the results of 100 Monte Carlo realizations starting from a flat distribution in both spin magnitudes and initial orientations for four representative cases (the alignment process and the evolution of the spin parameter are almost identical in all other runs).
The time evolution of the relative angle, $\theta$, between the spin of the SMBH and the orbital angular momentum of the circum-nuclear disk is also shown (central panel).
The inner (outer) shaded areas in the upper and central insets correspond to 1-sigma (2-sigma) deviations from the mean values, over 100 random initial spin realizations\footnote{
We checked that changing the number of initial realizations the final results are not affected significantly.
}.
The corresponding luminosity in units of the Eddington luminosity, $f_{\rm Edd}$, is plotted, as well (bottom panel).
All the results discussed in the following, where not specified otherwise, are obtained computing the properties of the accreting material at a distance from the central SMBH of 3~pc, i.e. three softening lengths of the low resolution runs, and for star formation timescale of $\tsfr = 1\,\rm Myr$.
Top-left panels refer to the reference run with axial wind and radiative cooling down to $\sim 10^4\,\rm K$, top-right panels to the corresponding run with low-temperature cooling (below $10^4\,\rm K$), bottom-left panels to the run with isotropic winds and cooling down to $\sim 10^4\,\rm K$, and bottom-right panels to the reference run including RT.
\\
In general, SMBH spins rapidly loose memory of their initial orientation, and accretion torques suffice to align the spins with the angular momentum of the disk orbit on a short timescale ($\simlt 1-2$ Myr).
In all the cases, ${\mathbf J}_{\rm BH}$ tends to align with the angular momentum of the accreting gas, as extracted by the simulations and highlighted by the black lines in the upper and central insets, within a few Myr, regardless the initial orientation and magnitude of the SMBH spin, even if initially completely anti-aligned.
After the first, fast alignment of the spin with the orbital angular momentum of the circum-nuclear disk, the accretion flow is mostly coherent, i.e. the inflow traces the angular momentum of the large-scale  circum-nuclear disk that feeds the SMBH, and the gas luminosity near the center is very
close to the Eddington luminosity.
Only in the RT runs (bottom-right panel) we find inclination angles that are slightly larger: this is due to the more realistic treatment of entropy injection by radiative sources, that is powered by radiation heating (missing in the other cases).
This does not change our overall conclusions, because typical
values are still $< ~50^{\circ}$, and only rarely isolated spikes are found.
After the ``fast'' spin alignment, accretion proceeds essentially coherently, i.e. the gas accretes on planes that have an inclination angle with respect to ${\mathbf J}_{\rm BH}$ that is significantly smaller than $90^{\circ}$.
As a consequence, $a$ monotonically increases with time, except for $a$ very close to unity, where it stays roughly constant, or decreases slightly.
Since for $a\approx 0$ the alignment timescale ($\propto a^{5/7}$) is extremely short \cite[][]{Perego2009}, slowly rotating SMBHs start immediately to accrete in a prograde fashion, and the lower boundary of the shaded areas are not expected to decrease with time on timescales of $\sim \rm Myr$.
Once a BH starts accreting from a prograde disk, the spin magnitude also increases over short timescales, close to the maximum allowed one.
\\
The resulting accretion rate, $f_{\rm Edd}$, shown in the lower insets, is more seriously affected by the different physical mechanisms
By comparing the EVP1e2-SFRT1 and the EVP1e2-SFRT1-LowT runs (upper right panel)
it emerges that low-temperature cooling contributes to a slight enhancement of $f_{\rm Edd}$.
More quantitatively, the average increment during the first 15~Myr is roughly $\sim 4$ per cent,
with temporary peaks of a factor of $\sim 2$.
In fact, at $\sim 5\,\rm Myr$, $f_{\rm Edd}$ results to be $\sim 0.3$ for the standard run, and $\sim 0.6$ for the low-temperature case.
This is related to the more efficient gas condensation process that can take place in such conditions.
In this case, gas can loose more pressure support, reach lower temperatures, and form smaller clumps with very high densities on short timescales.
Both formation of dense structures and gas compression from SN shocks reshuffle the local angular momentum of the disk strengthening inflows onto the central black hole \cite[][]{wada2009, kawakatu2009, hobbs2011}.
The overall effect is limited on short timescales, since in the long run (e.g. $\gtrsim 15\,\rm Myr$) the trends tend to converge mainly because of gas consumption, as shown by the decrement in the star formation activity.
\\
When comparing isotropic winds to axial winds (see upper-left and lower-left panels), we notice an initial drop in accretion rate for the isotropic case.
The very first episodes of star formation happen close to the black hole, where the gas density reaches initially its peak.  Because of the disk-like geometry of our gas distribution, isotropic winds can interact with close gas particle more efficiently, initially evacuating the central region of the disk, and hence decreasing the accretion rate.
However, as soon as star formation picks up in the outer regions of the disk, the accretion rate onto the black hole raises again, up to values even slightly higher than in the runs with axial winds.
This is because isotropic winds mix the angular momentum of gas more efficiently.
In the RT case (bottom-right panel), heating from radiative feedback keeps the gas hotter, strongly suppressing the formation of dense substructures and star formation.
As a consequence, $f_{\rm Edd}$ is decreased by a factor of a few, with peak values of $f_{\rm Edd} \lesssim 1/2$.
\\
In our highest resolution run (figure~\ref{fig:hires}), we can resolve almost one star at a time. Even in this run we find results similar to the lower resolution runs, with the only difference that we can follow gas inflow down to sub-parsec scales (0.3~pc), deeper in the potential well of the central SMBH, where the accretion rate is, in average, a factor of $\sim 2-3$ larger.
\\
We emphasize that these results apply only to quasars in merger remnants, not to those fed by disk instabilities and cold streams. In those cases, massive clumps may chaoticize accretion \cite[e.g.][]{Bournaud_et_al_2011, Dubois2012}, since they would usually infall from large scales carrying their own angular momentum.

\subsection{Parameter study}

\begin{figure*}
  \centering
  \includegraphics[width=0.32\textwidth]{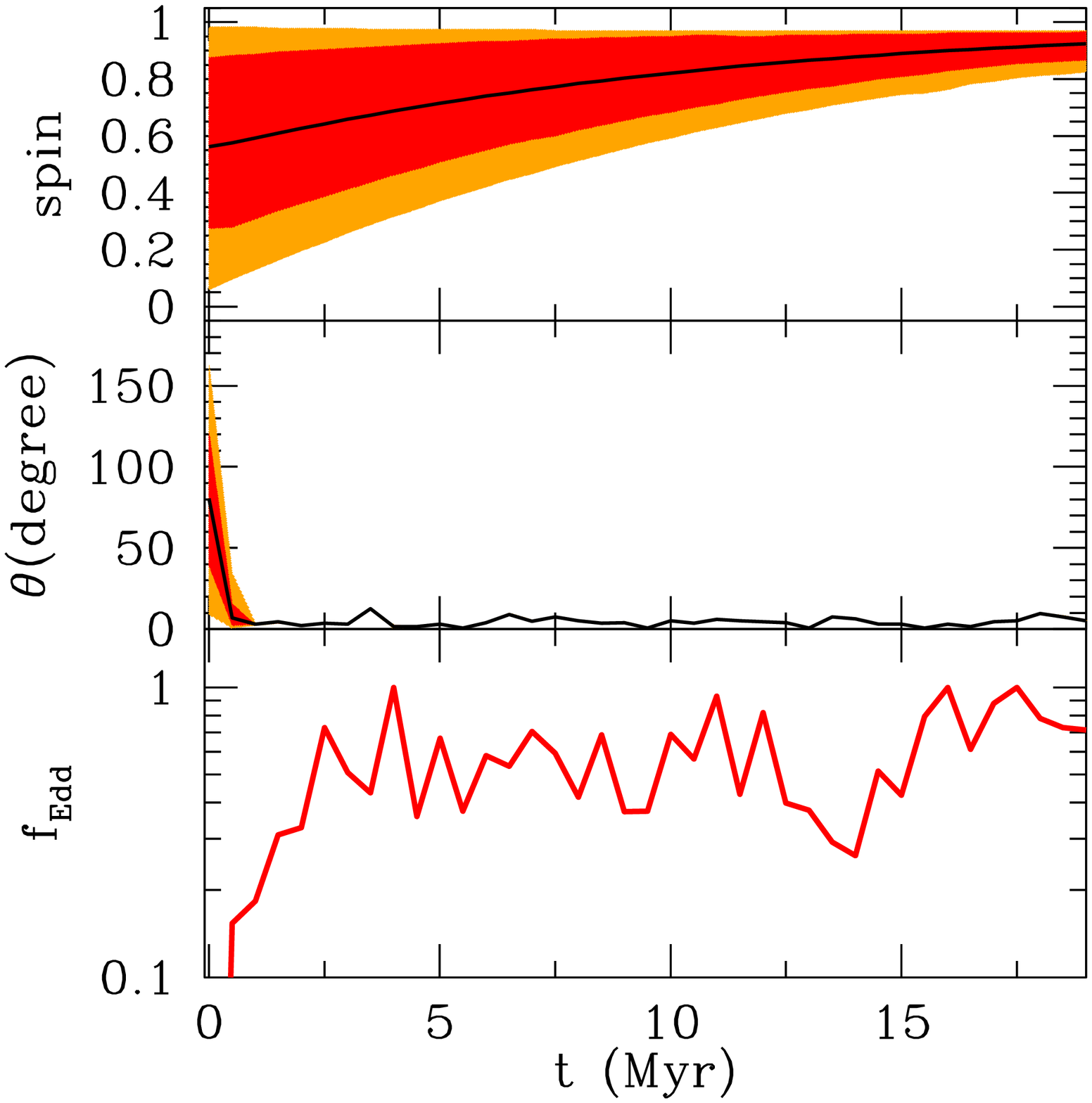}
  \includegraphics[width=0.32\textwidth]{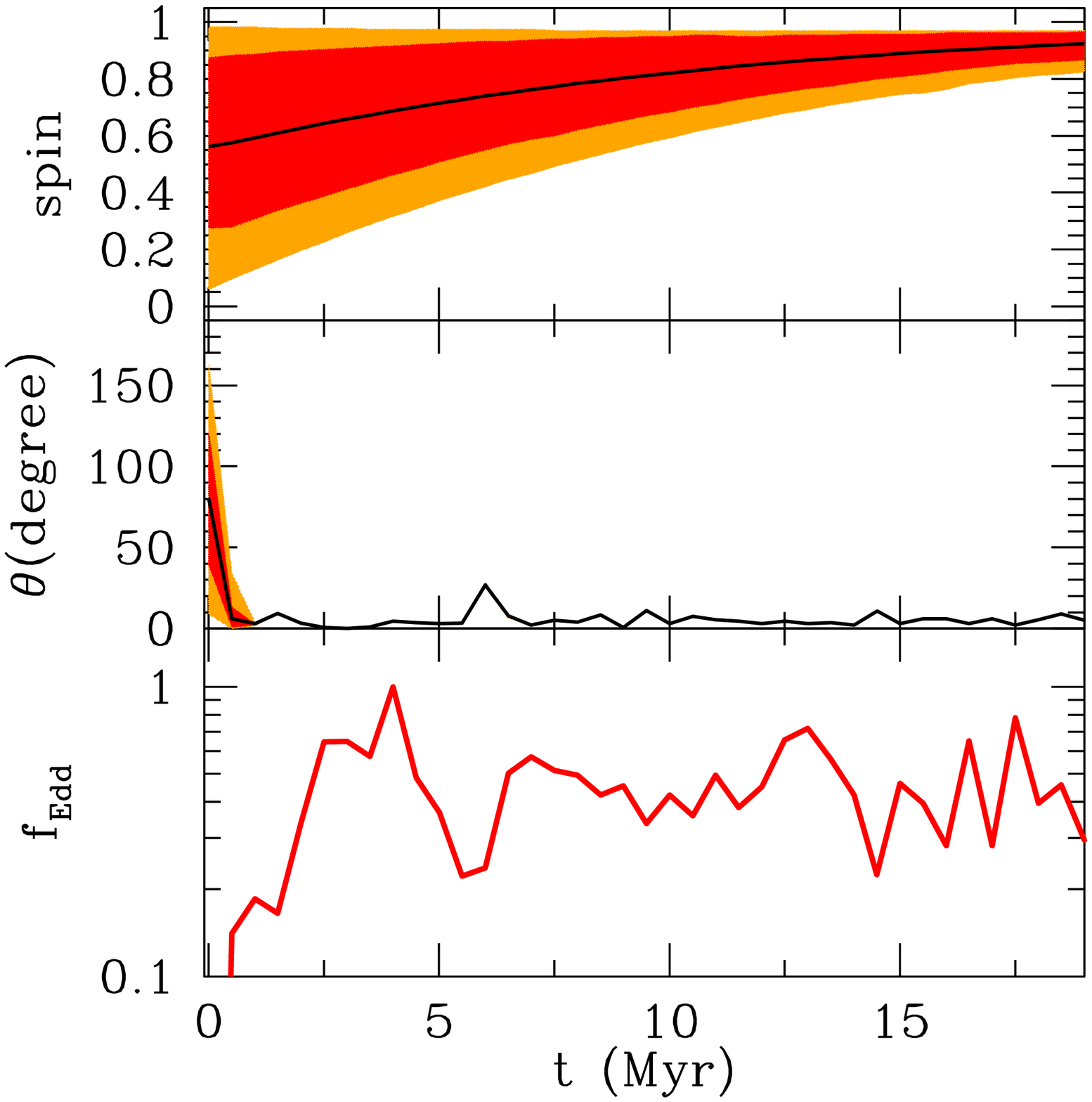}
  \includegraphics[width=0.32\textwidth]{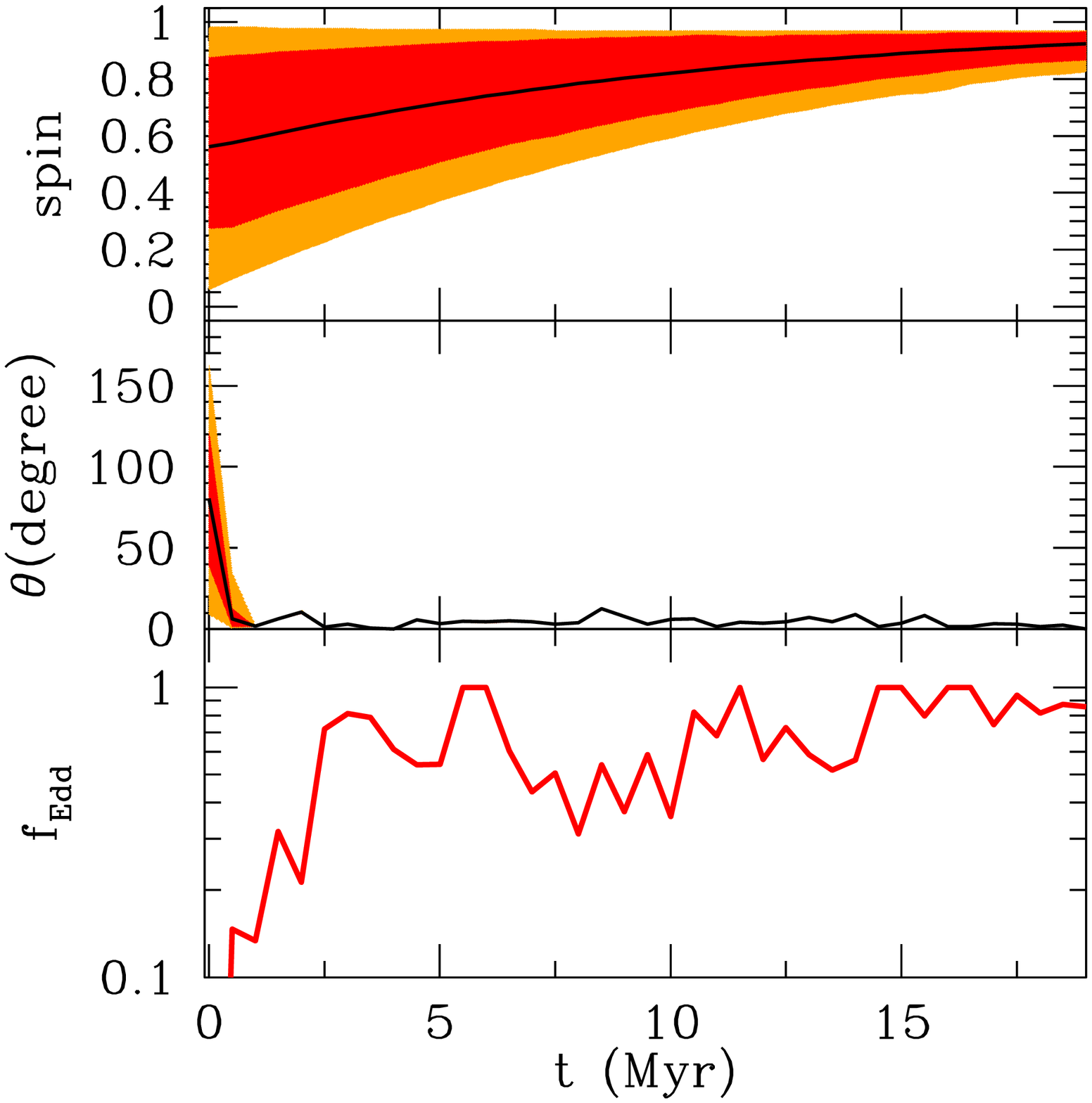}
  \caption{\small
    As in figure~\ref{fig:spins} for runs performed assuming different star formation timescales: $\tsfr = $~50 (EVP1e2-SFRT50, left panel), 100 (EVP1e2-SFRT100, middle panel), and 300~Myr (EVP1e2-SFRT300, right panel).
  }
  \label{fig:tsfr}
\end{figure*}

\begin{figure*}
  \centering
  \includegraphics[width=0.49\textwidth]{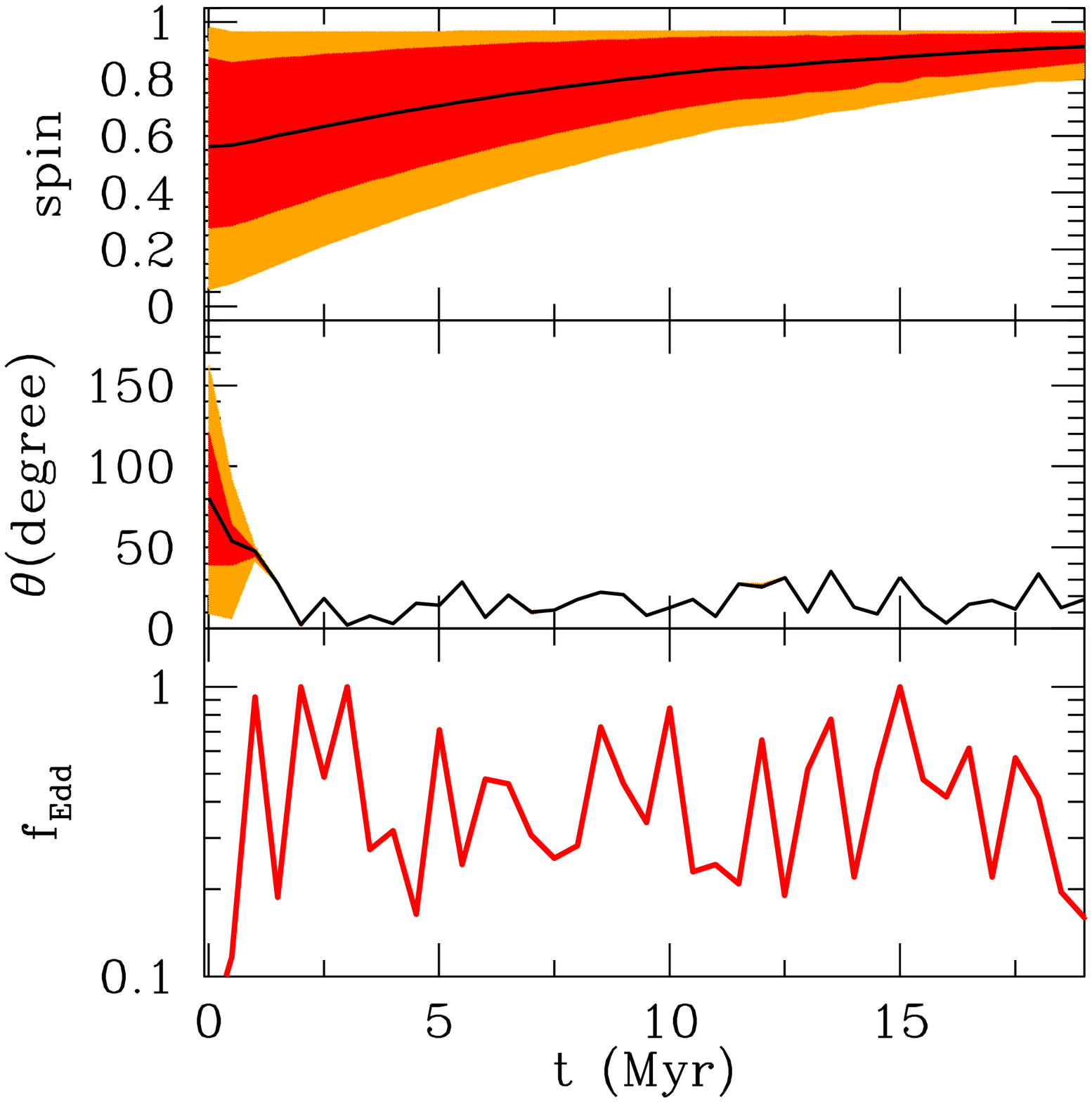}
  \includegraphics[width=0.49\textwidth]{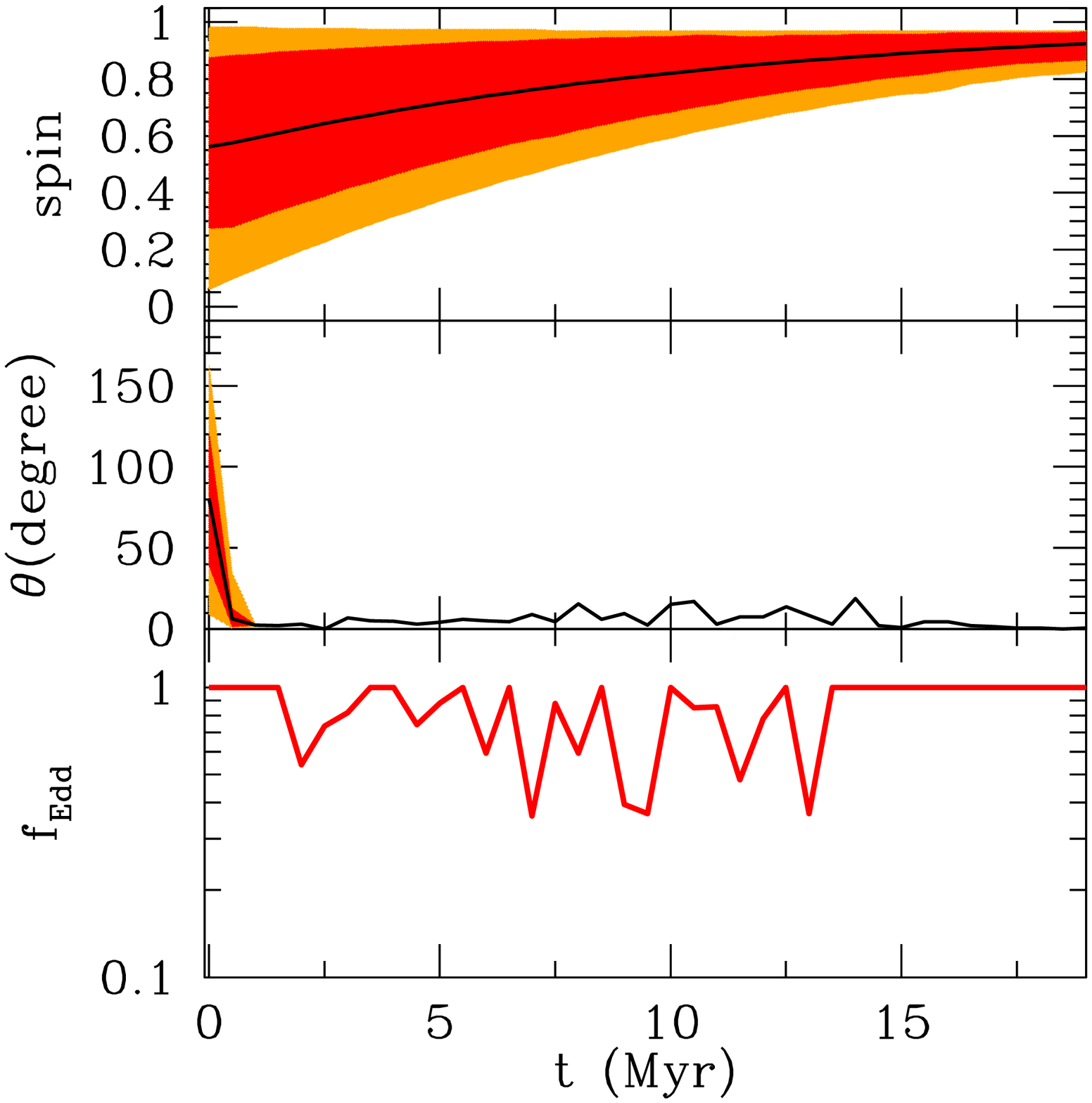}
  \caption{\small
    As in figure~\ref{fig:spins} for runs assuming star formation timescales of 1 (left) and 300 (right) Myr, both with full stellar evolution and metal enrichment calculations (SE). For a direct comparison with the corresponding $Z=0$ cases, see top-left plot in figure~\ref{fig:spins}, and right plot in figure~\ref{fig:tsfr}, respectively.
  }
  \label{fig:Z}
\end{figure*}

\noindent
To draw more robust and definitive conclusions, we finish by showing the main findings arising from a parameter study for different cases.
In particular, to check the dependencies on star formation we explore the changes induced by adopting different star formation timescales, gas density thresholds, evaporation factors, self-consistent stellar evolution, and metal spreading.
\\
Similarly to figure~\ref{fig:spins} and \ref{fig:hires}, where we explored the case $\tsfr = 1\,\rm Myr$, in figure~\ref{fig:tsfr} we display the time evolution of the spin parameter, of the spin orientation angle, and of the luminosity in Eddington units, $f_{\rm Edd}$, also for the runs performed with $\tsfr = $~50, 100, 300~Myr (in the cases with $\tsfr = 1-20\,\rm Myr$ no relevant differences are found, hence we will not focus on each of them). The general trends show that, despite the good agreements of the spin parameters (in all the cases ranging between $\sim 0.6$ and 0.9 at $t \sim 10\,\rm Myr$) and of the inclination angles (always quite small), models with larger star formation timescales (i.e. smaller efficiency), as the one with $\tsfr = 300\,\rm Myr$, can nevertheless accrete slightly more mass and sustain $f_{\rm Edd}$ at slightly larger values ($f_{\rm Edd}\sim 0.8-1$ at $\gtrsim 15\,\rm Myr$) than models with more efficient star formation and shorter $\tsfr$.
We also notice that, in models with larger $\tsfr$, the formation of long-lived gas clumps that could stir the medium seems not to be very efficient in increasing the chaotic behaviour of the CND, with $f_{\rm Edd}$ reaching roughly unity after $\gtrsim 10-15~\rm Myr$.
\\
Our conclusions could be affected by the ability of resolving properly the cold, dense, star forming regions. As a check, we analyze the cases carried out by assuming different gas density thresholds for star formation in the range $10^2 - 10^8\,\rm cm^{-3}$ (see supplementary online material).
%
%
It emerges that models with low-density thresholds have a more chaotic behaviour and show larger spreads in the results.
Indeed, for the scenarios with a threshold of $\sim 10^2-10^3\,\rm cm^{-3}$, the spin parameter is roughly $a\sim 0.4 - 0.95$ and $\theta$ can smoothly increase above $\sim 30^o$, or so.
The more chaotic behavior is visible mainly at initial times, when first bursts of star formation take place, and the spreads in the data are wider than in the higher-density-threshold cases.
Thus, the accretion rate is easily inhibited (with $f_{\rm Edd}$ dropping below $\sim 0.1$), because even material at relatively low-densities experiences star formation and gets evacuated by the consequent thermal heating and feedback processes.
Larger, more realistic, density thresholds, instead, allow the gas of the CND to cool and actually condense, so that star formation takes place only in very clumpy regions on the disk.
Dense, cold fragments are affected in a minor way by feedback processes, can survive them, and continue accreting on the central BH.
As a consequence, the CND can keep a more stable and ordered shape, and the cases with a threshold $\gtrsim 10^5\,\rm cm^{-3}$ converge rapidly (in $\lesssim 1\,\rm Myr$) to similar trends.
Moreover, they are compatible with the fiducial value of $\sim 10^4\,\rm cm^{-3}$ used throughout this paper.
\\
The effects of changes in the evaporation factor (see supplementary online material)
%
%
lead to similar spin evolution, irrespectively from the detailed values used, and to inclination angles always below $\sim 50^o$.
In general, despite the non-very-regular trends, some boosts in $\theta$ are present when there is strong star formation activity (figure~\ref{fig:sfr}) for all the cases, with corresponding decrements in the Eddington fraction, since hot gas gets more pressure supported and provokes smaller inflows towards the central regions. Instead, when star formation rates drop down, the resulting $\theta$ values decrease and the $f_{\rm Edd}$ values get closer to unity.
This variability is a consequence of the interplay between star formation heating and gas cooling that regulate the behaviour of the environment.
Stellar feedback heats the medium, while simultaneous cooling processes cool down the gas, stabilize the disk, and trigger the next star formation episodes.
\\
Complete full stellar evolution and metal spreading, displayed in figure~\ref{fig:Z}, does not imply dramatic changes in the spin results.
The sharper variations (on timescales of $\sim 1 \,\rm Myr$) of the inclination angle and of $f_{\rm Edd}$ are due to the larger radiative rates in presence of metals, and, hence, faster cooling and star formation processes, that respectively boost and inhibit gas inflows.
This is mostly evident for $ \tsfr=1 \,\rm Myr$, where metal pollution takes place earlier than in the case of longer $\tsfr=300 \,\rm Myr$.
In the former case, the effects of short star formation timescales (see also right panel in figure~\ref{fig:sfr2}) are enhanced by the stronger radiative losses of the newly enriched material, that quickly cool the ambient medium and temporarily enhance gas infall.
In the latter case, the star formation rate at the same times is a factor of a few to $\sim 10$ more limited, and cooling from metals is able to efficiently stabilize the disk against the chaotic motions from the rarer stellar feedback.
These conclusions on the stabilizing role of metals are easily cross-checked by a direct comparison with the corresponding $Z=0$ cases, in the top-left plot of figure~\ref{fig:spins}, and in the right plot of figure~\ref{fig:tsfr}, for $\tsfr=1~\rm Myr$ and $\tsfr=300~\rm Myr$, respectively.


\section{Discussion and conclusions}\label{sect:conclusions}


\noindent
We have presented a number of thre-dimensional, N-body, hydrodynamical, chemistry simulations of the behaviour of circum-nuclear discs around super-massive black holes.
We considered the relevant local physical processes that could alter the coherency of gaseous flows onto the central regions, i.e.: cooling, star formation, feedback effects, and radiative transfer from stellar sources.
Our goal was to explore the implications of these several phenomena on the stability of circum-nuclear discs around SMBHs and the consequent effects on the black-hole spin evolution.
Our investigation considers mainly those ``local'' processes that could increase the turbulence of the gas (such as the gas self-gravity, cooling, star-formation, and feedback from young stars), while global processes, such as the precession of the nuclear gas structures in a less symmetric, tri-axial potential, are not taken in account, as we started with an axisymmetric distribution of gas and stars. Global instabilities or asymmetries that could decrease the degree of coherency in the BH fuelling \cite[as discussed e.g. in][]{Hopkins2011} are beyond the aims of this work.
\\
Our study does not include AGN feedback. The energy liberated by an AGN may be by far larger than the typical energies delivered through stellar feedback in the central regions of the galaxy.
The effectiveness of the AGN in altering the CND gas dynamics, however, depends on the jet geometry and on the fraction of energy actually deposited in the nuclear regions.
In general terms, we can make some inferences on the expected impact of AGN feedback from the results of (two-dimensional) dedicated simulations such as those by \cite{Novak2011}.
They find that mechanical feedback (winds) dominates the effect on the local gas dynamics, and that this effect is stronger when the wind is confined in a cone. Nevertheless, even during a high-accretion event a beamed wind does not strongly affect the gas located perpendicularly to the outflow.
These simulations differ from our three-dimensional setup, though. On the one hand, they do not include any dense and rotationally supported gas structure, that would reduce even further the effect of the AGN feedback. On the other hand, they keep the orientation of the conical outflow constant. If the orientation would follow the rearrangement of the SMBH spin, it would in some cases transiently impinge on the CND, modifying more strongly its dynamics and the properties of the accreting flow.
Definitely, more self-consistent investigations are needed to assess the interplay between AGN feedback and CND dynamics, but this first study, focusing on the dynamical perturbations self-generated by the CND, will facilitate distinguishing the effects of AGN feedback in future works.
\\
Summarizing, our analysis suggests that in most of the cases local star formation episodes and feedback effects are able to partially break the general coherency of the gas flow, but do not inject enough energy and momentum into the gas near SMBHs to create a strongly chaotic environment that completely randomizes the orbits of the inflowing particles. As a conclusion, we find that initially maximally rotating SMBHs are slightly spun down, and initially slow-rotating black holes are spun up, leading to upper-intermediate equilibrium values, $a\simeq 0.6-0.9$ (corresponding to radiative efficiencies $\epsilon \simeq 9\%-15\%$).
\\
Considering the whole parameter space, it results that different feedback mechanisms can affect the CND gas dynamics and coherency in different ways.
Mechanical (kinetic and thermal) feedback processes are responsible for causing temporary, moderate, changes in the inclination angle and Eddington fraction.
Metal feedback, instead, introduces sharper variations, but has a generally stabilizing role for the CND gas flows, since it facilitates the formation of cold clumps.
Radiative feedback is responsible for inducing a more chaotic regime with increments of the alignment angle of at least a factor of a few (up to $\sim 50^o$), and for limiting the accretion capabilities of the central BH of a factor $\gtrsim 2$.
\\
We note that our results are based on the angular momentum of the material that is effectively accreted on the SMBH, and not on the global alignment of the nuclear disc with the large scale properties of the host galaxy \cite[][]{Hopkins2011}.
While the nuclear disc can be misaligned with respect to the galaxy disc, the relevant quantity to address spin evolution is the direction of the angular momentum of the material within the nuclear disc that eventually feeds the SMBH, not the alignment or misalignment between nuclear disc and the host.
These results are in line with the suggestion that jets are preferentially perpendicular to dust lanes \cite[][]{Verdoes2005}, which are believed to be the remnants of galaxy mergers -- a low-mass analogue of nuclear discs. Further hints for at least partial alignment are obtained from analyses of active galactic nuclei (AGN) and their hosts in the Sloan survey \cite[][]{Lagos2011}.  According to these studies, the material feeding the black hole is expected to accrete in a somewhat coherent fashion.
\\
Observationally, spin magnitudes of SMBHs powering AGNs can be estimated directly by measuring the width of the K$_\alpha$ iron line, at 6.4~keV, through X-ray spectroscopy: recent determinations with
{\it Chandra}\footnote{http://chandra.si.edu/},
{\it XMM-Newton}\footnote{http://xmm.esac.esa.int/}, 
and {\it Suzaku}\footnote{http://www.isas.jaxa.jp/e/enterp/missions/suzaku/}
have revealed variable, relativistic iron emission lines from the inner disc in many Seyfert galaxies \cite[][]{Brenneman2011}\footnote{ This iron line originates near the black hole, and it is distorted by relativistic effects \cite[][]{Fabian1989,Laor1991}.
In particular, the width of the line is much broader in the rotating case, because the last stable orbit is closer in, and the gravitational redshift is stronger.
}.
It is suggestive that all SMBHs with spin measured directly through X-ray spectroscopy have spin parameters between  $\sim$~0.6 and 1 \cite[][]{Brenneman2011}, in broad agreement with our results. Indeed, if after the quasar phase the mass of a SMBH has not changed much through gas accretion or SMBH mergers, the expectation is that today's spins are similar to the final spin gained during the last growth spurt.


\section*{Acknowledgments}

\noindent
We acknowledge V.~Springel for providing access to Gadget-3 code. U.M. acknowledges financial contribution from the Project HPC-Europa2, funded under the European Seventh Framework Programme, Infrastructure, Grant Agreement n. 228398, and kind hospitality at the Italian computing center (CINECA); he also acknowedges funding from the European Commission Seventh Framework Programme (FP7/2007-2013) Grant Agreement n. 267251.
M.P. research is part of the project GLENCO, funded under the European Seventh Framework Programme, Ideas, Grant Agreement n. 259349. 
M.V. acknowledges support from NASA, through award ATP NNX10AC84G; from NSF, through award AST 1107675; and the European Seventh Programme FP7/PEOPLE/2012-CIG (PCIG-GA-2011-303609). For the bibliographic research we made use of the tools offered by the NASA Astrophysics Data System (ADS) and by the JSTOR archive.


\bibliography{bibl}

\begin{thebibliography}{115}
\expandafter\ifx\csname natexlab\endcsname\relax\def\natexlab#1{#1}\fi

\bibitem[{{Abramowicz} {et~al.}(1978){Abramowicz}, {Jaroszynski}, \&
  {Sikora}}]{Abramowicz1978}
{Abramowicz}, M., {Jaroszynski}, M., \& {Sikora}, M. 1978, \aap, 63, 221

\bibitem[{{Aguirre} {et~al.}(2001){Aguirre}, {Hernquist}, {Schaye}, {Katz},
  {Weinberg}, \& {Gardner}}]{Aguirre_et_al_2001}
{Aguirre}, A., {Hernquist}, L., {Schaye}, J., {Katz}, N., {Weinberg}, D.~H., \&
  {Gardner}, J. 2001, \apj, 561, 521

\bibitem[{{Bambi} \& {Barausse}(2011)}]{BambiBarausse2011}
{Bambi}, C. \& {Barausse}, E. 2011, \prd, 84, 084034

\bibitem[{{Bardeen}(1970)}]{Bardeen1970}
{Bardeen}, J.~M. 1970, Nature, 226, 64

\bibitem[{{Bardeen} \& {Petterson}(1975)}]{bardeenpetterson1975}
{Bardeen}, J.~M. \& {Petterson}, J.~A. 1975, ApJL, 195, L65+

\bibitem[{{Bardeen} {et~al.}(1972){Bardeen}, {Press}, \&
  {Teukolsky}}]{Bardeenetal1972}
{Bardeen}, J.~M., {Press}, W.~H., \& {Teukolsky}, S.~A. 1972, \apj, 178, 347

\bibitem[{{Bate} {et~al.}(2010){Bate}, {Lodato}, \&
  {Pringle}}]{BateLodatoPringle2010}
{Bate}, M.~R., {Lodato}, G., \& {Pringle}, J.~E. 2010, \mnras, 401, 1505

\bibitem[{{Blandford} \& {Payne}(1982)}]{BlandfordPayne1982}
{Blandford}, R.~D. \& {Payne}, D.~G. 1982, \mnras, 199, 883

\bibitem[{{Blandford} \& {Znajek}(1977)}]{Blandford1977}
{Blandford}, R.~D. \& {Znajek}, R.~L. 1977, \mnras, 179, 433

\bibitem[{{Bogdanovi{\'c}} {et~al.}(2007){Bogdanovi{\'c}}, {Reynolds}, \&
  {Miller}}]{Bogdanovic2007}
{Bogdanovi{\'c}}, T., {Reynolds}, C.~S., \& {Miller}, M.~C. 2007, ApJL, 661,
  L147

\bibitem[{{Booth} \& {Schaye}(2009)}]{Booth2009}
{Booth}, C.~M. \& {Schaye}, J. 2009, \mnras, 398, 53

\bibitem[{{Bournaud} {et~al.}(2011){Bournaud}, {Dekel}, {Teyssier}, {Cacciato},
  {Daddi}, {Juneau}, \& {Shankar}}]{Bournaud_et_al_2011}
{Bournaud}, F., {Dekel}, A., {Teyssier}, R., {Cacciato}, M., {Daddi}, E.,
  {Juneau}, S., \& {Shankar}, F. 2011, \apjl, 741, L33

\bibitem[{{Bower} {et~al.}(2006){Bower}, {Benson}, {Malbon}, {Helly}, {Frenk},
  {Baugh}, {Cole}, \& {Lacey}}]{Bower2006}
{Bower}, R.~G., {Benson}, A.~J., {Malbon}, R., {Helly}, J.~C., {Frenk}, C.~S.,
  {Baugh}, C.~M., {Cole}, S., \& {Lacey}, C.~G. 2006, \mnras, 370, 645

\bibitem[{{Boyer} \& {Lindquist}(1967)}]{BoyerLindquist1967}
{Boyer}, R.~H. \& {Lindquist}, R.~W. 1967, Journal of Mathematical Physics, 8,
  265

\bibitem[{{Brenneman} {et~al.}(2011){Brenneman}, {Reynolds}, {Nowak}, {Reis},
  {Trippe}, {Fabian}, {Iwasawa}, {Lee}, {Miller}, {Mushotzky}, {Nandra}, \&
  {Volonteri}}]{Brenneman2011}
{Brenneman}, L.~W., {Reynolds}, C.~S., {Nowak}, M.~A., {Reis}, R.~C., {Trippe},
  M., {Fabian}, A.~C., {Iwasawa}, K., {Lee}, J.~C., {Miller}, J.~M.,
  {Mushotzky}, R.~F., {Nandra}, K., \& {Volonteri}, M. 2011, ArXiv e-prints

\bibitem[{{Burkert}(2006)}]{Burkert2006}
{Burkert}, A. 2006, Comptes Rendus Physique, 7, 433

\bibitem[{{Campanelli} {et~al.}(2007){Campanelli}, {Lousto}, {Zlochower}, \&
  {Merritt}}]{Campanelli2007}
{Campanelli}, M., {Lousto}, C.~O., {Zlochower}, Y., \& {Merritt}, D. 2007,
  Physical Review Letters, 98, 231102

\bibitem[{{Carter}(1968)}]{Carter1968}
{Carter}, B. 1968, Physical Review, 174, 1559

\bibitem[{{Cen} \& {Ostriker}(1992{\natexlab{a}})}]{Cen1992}
{Cen}, R. \& {Ostriker}, J.~P. 1992{\natexlab{a}}, ApJL, 399, L113

\bibitem[{{Cen} \& {Ostriker}(1992{\natexlab{b}})}]{CenOstriker1992}
---. 1992{\natexlab{b}}, \apjl, 399, L113

\bibitem[{{Croton} {et~al.}(2006){Croton}, {Springel}, {White}, {De Lucia},
  {Frenk}, {Gao}, {Jenkins}, {Kauffmann}, {Navarro}, \& {Yoshida}}]{Croton2006}
{Croton}, D.~J., {Springel}, V., {White}, S.~D.~M., {De Lucia}, G., {Frenk},
  C.~S., {Gao}, L., {Jenkins}, A., {Kauffmann}, G., {Navarro}, J.~F., \&
  {Yoshida}, N. 2006, MNRAS, 365, 11

\bibitem[{{Davies} {et~al.}(2003){Davies}, {Tacconi}, {Genzel}, \&
  {Thatte}}]{Davies2003}
{Davies}, R., {Tacconi}, L., {Genzel}, R., \& {Thatte}, N. 2003, ArXiv
  Astrophysics e-prints

\bibitem[{{Davies} {et~al.}(2004){Davies}, {Tacconi}, \& {Genzel}}]{Davies2004}
{Davies}, R.~I., {Tacconi}, L.~J., \& {Genzel}, R. 2004, \apj, 602, 148

\bibitem[{{de Vaucouleurs}(1948)}]{deVaucouleurs1948}
{de Vaucouleurs}, G. 1948, Annales d'Astrophysique, 11, 247

\bibitem[{{Di Matteo} {et~al.}(2008){Di Matteo}, {Colberg}, {Springel},
  {Hernquist}, \& {Sijacki}}]{Dimatteo2008}
{Di Matteo}, T., {Colberg}, J., {Springel}, V., {Hernquist}, L., \& {Sijacki},
  D. 2008, ApJ, 676, 33

\bibitem[{{Di Matteo} {et~al.}(2005){Di Matteo}, {Springel}, \&
  {Hernquist}}]{DiMatteo2005}
{Di Matteo}, T., {Springel}, V., \& {Hernquist}, L. 2005, \nat, 433, 604

\bibitem[{{Dotti} {et~al.}(2010){Dotti}, {Volonteri}, {Perego}, {Colpi},
  {Ruszkowski}, \& {Haardt}}]{Dotti2010}
{Dotti}, M., {Volonteri}, M., {Perego}, A., {Colpi}, M., {Ruszkowski}, M., \&
  {Haardt}, F. 2010, MNRAS, 402, 682

\bibitem[{{Downes} \& {Eckart}(2007)}]{DownesEckart2007}
{Downes}, D. \& {Eckart}, A. 2007, \aap, 468, L57

\bibitem[{{Downes} \& {Solomon}(1998)}]{Downes1998}
{Downes}, D. \& {Solomon}, P.~M. 1998, \apj, 507, 615

\bibitem[{{Dubois} {et~al.}(2012){Dubois}, {Pichon}, {Haehnelt}, {Kimm},
  {Slyz}, {Devriendt}, \& {Pogosyan}}]{Dubois2012}
{Dubois}, Y., {Pichon}, C., {Haehnelt}, M., {Kimm}, T., {Slyz}, A.,
  {Devriendt}, J., \& {Pogosyan}, D. 2012, \mnras, 423, 3616

\bibitem[{{Evans} {et~al.}(2006){Evans}, {Solomon}, {Tacconi}, {Vavilkin}, \&
  {Downes}}]{Evans2006}
{Evans}, A.~S., {Solomon}, P.~M., {Tacconi}, L.~J., {Vavilkin}, T., \&
  {Downes}, D. 2006, \aj, 132, 2398

\bibitem[{{Fabian} {et~al.}(1989){Fabian}, {Rees}, {Stella}, \&
  {White}}]{Fabian1989}
{Fabian}, A.~C., {Rees}, M.~J., {Stella}, L., \& {White}, N.~E. 1989, MNRAS,
  238, 729

\bibitem[{{Fabjan} {et~al.}(2010){Fabjan}, {Borgani}, {Tornatore}, {Saro},
  {Murante}, \& {Dolag}}]{Fabian2010}
{Fabjan}, D., {Borgani}, S., {Tornatore}, L., {Saro}, A., {Murante}, G., \&
  {Dolag}, K. 2010, \mnras, 401, 1670

\bibitem[{{Felice}(1968)}]{Felice1968}
{Felice}, F. 1968, Nuovo Cimento B Serie, 57, 351

\bibitem[{{Gammie} {et~al.}(2004){Gammie}, {Shapiro}, \&
  {McKinney}}]{Gammie2004}
{Gammie}, C.~F., {Shapiro}, S.~L., \& {McKinney}, J.~C. 2004, ApJ, 602, 312

\bibitem[{{Ghosh} \& {Abramowicz}(1997)}]{GhoshAbramowicz1997}
{Ghosh}, P. \& {Abramowicz}, M.~A. 1997, \mnras, 292, 887

\bibitem[{{Haehnelt} {et~al.}(1998){Haehnelt}, {Natarajan}, \&
  {Rees}}]{HNR1998}
{Haehnelt}, M.~G., {Natarajan}, P., \& {Rees}, M.~J. 1998, MNRAS, 300, 817

\bibitem[{{Hartmann} {et~al.}(2001){Hartmann}, {Ballesteros-Paredes}, \&
  {Bergin}}]{Hartmann2001}
{Hartmann}, L., {Ballesteros-Paredes}, J., \& {Bergin}, E.~A. 2001, \apj, 562,
  852

\bibitem[{{Hernquist}(1990)}]{Hernquist1990}
{Hernquist}, L. 1990, \apj, 356, 359

\bibitem[{{Hernquist} \& {Springel}(2003)}]{Hernquist2003}
{Hernquist}, L. \& {Springel}, V. 2003, \mnras, 341, 1253

\bibitem[{{Hobbs} {et~al.}(2011){Hobbs}, {Nayakshin}, {Power}, \&
  {King}}]{hobbs2011}
{Hobbs}, A., {Nayakshin}, S., {Power}, C., \& {King}, A. 2011, \mnras, 413,
  2633

\bibitem[{{Hopkins} {et~al.}(2011){Hopkins}, {Hernquist}, {Hayward}, \&
  {Narayanan}}]{Hopkins2011}
{Hopkins}, P.~F., {Hernquist}, L., {Hayward}, C.~C., \& {Narayanan}, D. 2011,
  ArXiv e-prints

\bibitem[{{Hopkins} {et~al.}(2005){Hopkins}, {Hernquist}, {Martini}, {Cox},
  {Robertson}, {Di Matteo}, \& {Springel}}]{hopkins2005}
{Hopkins}, P.~F., {Hernquist}, L., {Martini}, P., {Cox}, T.~J., {Robertson},
  B., {Di Matteo}, T., \& {Springel}, V. 2005, {ApJ}, 625, L71

\bibitem[{{Hopkins} \& {Quataert}(2010)}]{hopkinsquataert2010}
{Hopkins}, P.~F. \& {Quataert}, E. 2010, \mnras, 407, 1529

\bibitem[{{Katz}(1992)}]{Katz1992}
{Katz}, N. 1992, \apj, 391, 502

\bibitem[{{Katz} \& {Gunn}(1991)}]{KatzGunn1991}
{Katz}, N. \& {Gunn}, J.~E. 1991, \apj, 377, 365

\bibitem[{{Katz} {et~al.}(1992){Katz}, {Hernquist}, \&
  {Weinberg}}]{Katz_et_al_1992}
{Katz}, N., {Hernquist}, L., \& {Weinberg}, D.~H. 1992, \apjl, 399, L109

\bibitem[{{Katz} {et~al.}(1996){Katz}, {Weinberg}, \&
  {Hernquist}}]{Katz_et_al_1996}
{Katz}, N., {Weinberg}, D.~H., \& {Hernquist}, L. 1996, \apjs, 105, 19

\bibitem[{{Kawakatu} \& {Wada}(2009)}]{kawakatu2009}
{Kawakatu}, N. \& {Wada}, K. 2009, in Astronomical Society of the Pacific
  Conference Series, Vol. 408, The Starburst-AGN Connection, ed. {W.~Wang,
  Z.~Yang, Z.~Luo, \& Z.~Chen}, 148

\bibitem[{{Kennicutt}(1998)}]{Kennicutt1998}
{Kennicutt}, Jr., R.~C. 1998, \araa, 36, 189

\bibitem[{{Kerr}(1963)}]{Kerr1963}
{Kerr}, R.~P. 1963, Physical Review Letters, 11, 237

\bibitem[{{King} {et~al.}(2005){King}, {Lubow}, {Ogilvie}, \&
  {Pringle}}]{King2005}
{King}, A.~R., {Lubow}, S.~H., {Ogilvie}, G.~I., \& {Pringle}, J.~E. 2005,
  MNRAS, 363, 49

\bibitem[{{King} \& {Pringle}(2006)}]{King2006}
{King}, A.~R. \& {Pringle}, J.~E. 2006, MNRAS, 373, L90

\bibitem[{{King} \& {Pringle}(2007)}]{KingPringle2007}
---. 2007, MNRAS, 377, L25

\bibitem[{{King} {et~al.}(2008{\natexlab{a}}){King}, {Pringle}, \&
  {Hofmann}}]{KingPringleHofmann2008}
{King}, A.~R., {Pringle}, J.~E., \& {Hofmann}, J.~A. 2008{\natexlab{a}},
  \mnras, 385, 1621

\bibitem[{{King} {et~al.}(2008{\natexlab{b}}){King}, {Pringle}, \&
  {Hofmann}}]{King2008}
---. 2008{\natexlab{b}}, MNRAS, 385, 1621

\bibitem[{{Krumholz} {et~al.}(2011){Krumholz}, {Dekel}, \&
  {McKee}}]{Krumholz2011}
{Krumholz}, M.~R., {Dekel}, A., \& {McKee}, C.~F. 2011, ArXiv e-prints

\bibitem[{{Kumar} \& {Pringle}(1985)}]{Kumar1985}
{Kumar}, S. \& {Pringle}, J.~E. 1985, MNRAS, 213, 435

\bibitem[{{Lagos} {et~al.}(2011){Lagos}, {Padilla}, {Strauss}, {Cora}, \&
  {Hao}}]{Lagos2011}
{Lagos}, C.~D.~P., {Padilla}, N.~D., {Strauss}, M.~A., {Cora}, S.~A., \& {Hao},
  L. 2011, \mnras, 414, 2148

\bibitem[{{Laor}(1991)}]{Laor1991}
{Laor}, A. 1991, ApJ, 376, 90

\bibitem[{{Li} {et~al.}(2006){Li}, {Hernquist}, {Robertson}, {Cox}, {Hopkins},
  {Springel}, {Gao}, {Di Matteo}, {Zentner}, {Jenkins}, \& {Yoshida}}]{Li2006}
{Li}, Y., {Hernquist}, L., {Robertson}, B., {Cox}, T.~J., {Hopkins}, P.~F.,
  {Springel}, V., {Gao}, L., {Di Matteo}, T., {Zentner}, A.~R., {Jenkins}, A.,
  \& {Yoshida}, N. 2006, ArXiv Astrophysics e-prints

\bibitem[{{Livio} {et~al.}(1999){Livio}, {Ogilvie}, \& {Pringle}}]{Livio1999}
{Livio}, M., {Ogilvie}, G.~I., \& {Pringle}, J.~E. 1999, \apj, 512, 100

\bibitem[{{Lodato} \& {Pringle}(2007)}]{LodatoPringle2007}
{Lodato}, G. \& {Pringle}, J.~E. 2007, \mnras, 381, 1287

\bibitem[{{Lousto} {et~al.}(2012){Lousto}, {Zlochower}, {Dotti}, \&
  {Volonteri}}]{Lousto2012}
{Lousto}, C.~O., {Zlochower}, Y., {Dotti}, M., \& {Volonteri}, M. 2012, ArXiv
  e-prints

\bibitem[{{Lynden-Bell}(1969)}]{Lynden-Bell1969}
{Lynden-Bell}, D. 1969, \nat, 223, 690

\bibitem[{{MacDonald} \& {Thorne}(1982)}]{MacDonaldThorne1982}
{MacDonald}, D. \& {Thorne}, K.~S. 1982, \mnras, 198, 345

\bibitem[{{Magorrian} {et~al.}(1998)}]{Magorrian1998}
{Magorrian}, J. {et~al.} 1998, AJ, 115, 2285

\bibitem[{{Maio} {et~al.}(2010{\natexlab{a}}){Maio}, {Ciardi}, {Dolag},
  {Tornatore}, \& {Khochfar}}]{Maio2010}
{Maio}, U., {Ciardi}, B., {Dolag}, K., {Tornatore}, L., \& {Khochfar}, S.
  2010{\natexlab{a}}, \mnras, 407, 1003

\bibitem[{{Maio} {et~al.}(2007){Maio}, {Dolag}, {Ciardi}, \&
  {Tornatore}}]{Maio2007}
{Maio}, U., {Dolag}, K., {Ciardi}, B., \& {Tornatore}, L. 2007, \mnras, 379,
  963

\bibitem[{{Maio} {et~al.}(2010{\natexlab{b}}){Maio}, {Khochfar}, {Johnson}, \&
  {Ciardi}}]{Maio2010b}
{Maio}, U., {Khochfar}, S., {Johnson}, J.~L., \& {Ciardi}, B.
  2010{\natexlab{b}}, ArXiv e-prints

\bibitem[{{Maio} {et~al.}(2011){Maio}, {Khochfar}, {Johnson}, \&
  {Ciardi}}]{Maio2011}
---. 2011, \mnras, 414, 1145

\bibitem[{{Martin}(1999)}]{Martin1999}
{Martin}, C.~L. 1999, \apj, 513, 156

\bibitem[{{Mayer} {et~al.}(2007){Mayer}, {Kazantzidis}, {Madau}, {Colpi},
  {Quinn}, \& {Wadsley}}]{Mayer2007}
{Mayer}, L., {Kazantzidis}, S., {Madau}, P., {Colpi}, M., {Quinn}, T., \&
  {Wadsley}, J. 2007, Science, 316, 1874

\bibitem[{{McKinney}(2005)}]{McKinney2005}
{McKinney}, J.~C. 2005, \apjl, 630, L5

\bibitem[{{Moderski} \& {Sikora}(1996)}]{ModerskiSikora1996}
{Moderski}, R. \& {Sikora}, M. 1996, \mnras, 283, 854

\bibitem[{{Moderski} {et~al.}(1998){Moderski}, {Sikora}, \&
  {Lasota}}]{Moderski1998}
{Moderski}, R., {Sikora}, M., \& {Lasota}, J.-P. 1998, MNRAS, 301, 142

\bibitem[{{Natarajan} \& {Pringle}(1998)}]{NatarajanPringle1998}
{Natarajan}, P. \& {Pringle}, J.~E. 1998, ApJL, 506, L97

\bibitem[{{Novak} {et~al.}(2011){Novak}, {Ostriker}, \& {Ciotti}}]{Novak2011}
{Novak}, G.~S., {Ostriker}, J.~P., \& {Ciotti}, L. 2011, \apj, 737, 26

\bibitem[{{Novikov} \& {Thorne}(1973)}]{NovikovThorne1973}
{Novikov}, I.~D. \& {Thorne}, K.~S. 1973, in Black Holes (Les Astres Occlus),
  ed. {C.~Dewitt \& B.~S.~Dewitt}, 343--450

\bibitem[{{Padovani} \& {Matteucci}(1993)}]{PadovaniMatteucci1993}
{Padovani}, P. \& {Matteucci}, F. 1993, \apj, 416, 26

\bibitem[{{Palla} \& {Stahler}(2000)}]{PallaStahler2000}
{Palla}, F. \& {Stahler}, S.~W. 2000, \apj, 540, 255

\bibitem[{{Perego} {et~al.}(2009){Perego}, {Dotti}, {Colpi}, \&
  {Volonteri}}]{Perego2009}
{Perego}, A., {Dotti}, M., {Colpi}, M., \& {Volonteri}, M. 2009, MNRAS, 399,
  2249

\bibitem[{{Peters}(1964)}]{Peters1964}
{Peters}, P.~C. 1964, Physical Review, 136, 1224

\bibitem[{{Petkova} \& {Maio}(2012)}]{PetkovaMaio2011}
{Petkova}, M. \& {Maio}, U. 2012, \mnras, 422, 3067

\bibitem[{{Petkova} \& {Springel}(2009)}]{Petkova2009}
{Petkova}, M. \& {Springel}, V. 2009, \mnras, 396, 1383

\bibitem[{{Petkova} \& {Springel}(2011)}]{Petkova2011}
---. 2011, \mnras, 412, 935

\bibitem[{{Rees} {et~al.}(1982){Rees}, {Begelman}, {Blandford}, \&
  {Phinney}}]{Rees1982}
{Rees}, M.~J., {Begelman}, M.~C., {Blandford}, R.~D., \& {Phinney}, E.~S. 1982,
  \nat, 295, 17

\bibitem[{{Sanders} \& {Mirabel}(1996)}]{SandersMirabel1996}
{Sanders}, D.~B. \& {Mirabel}, I.~F. 1996, \araa, 34, 749

\bibitem[{{Schartmann} {et~al.}(2011){Schartmann}, {Krause}, \&
  {Burkert}}]{schartmann2011}
{Schartmann}, M., {Krause}, M., \& {Burkert}, A. 2011, \mnras, 415, 741

\bibitem[{{Schartmann} {et~al.}(2009){Schartmann}, {Meisenheimer}, {Klahr},
  {Camenzind}, {Wolf}, \& {Henning}}]{schartmann2009}
{Schartmann}, M., {Meisenheimer}, K., {Klahr}, H., {Camenzind}, M., {Wolf}, S.,
  \& {Henning}, T. 2009, MNRAS, 393, 759

\bibitem[{{Scheuer} \& {Feiler}(1996)}]{Scheuer1996}
{Scheuer}, P.~A.~G. \& {Feiler}, R. 1996, MNRAS, 282, 291

\bibitem[{{Shakura} \& {Sunyaev}(1973)}]{ShakuraSunyaev1973}
{Shakura}, N.~I. \& {Sunyaev}, R.~A. 1973, \aap, 24, 337

\bibitem[{{Shakura} \& {Sunyaev}(1976)}]{ShakuraSunyaev1976}
---. 1976, \mnras, 175, 613

\bibitem[{{Shankar} {et~al.}(2009){Shankar}, {Weinberg}, \&
  {Miralda-Escud{\'e}}}]{Shankar2009}
{Shankar}, F., {Weinberg}, D.~H., \& {Miralda-Escud{\'e}}, J. 2009, \apj, 690,
  20

\bibitem[{{Shankar} {et~al.}(2011){Shankar}, {Weinberg}, \&
  {Miralda-Escude'}}]{Shankar2011arXiv}
{Shankar}, F., {Weinberg}, D.~H., \& {Miralda-Escude'}, J. 2011, ArXiv e-prints

\bibitem[{{Sijacki} {et~al.}(2007){Sijacki}, {Springel}, {di Matteo}, \&
  {Hernquist}}]{Sijacki2007}
{Sijacki}, D., {Springel}, V., {di Matteo}, T., \& {Hernquist}, L. 2007, MNRAS,
  380, 877

\bibitem[{{Sijacki} {et~al.}(2009){Sijacki}, {Springel}, \&
  {Haehnelt}}]{Sijacki2009}
{Sijacki}, D., {Springel}, V., \& {Haehnelt}, M.~G. 2009, MNRAS, 400, 100

\bibitem[{{Silk} \& {Rees}(1998)}]{Silk1998}
{Silk}, J. \& {Rees}, M.~J. 1998, {A\&A}, 331, L1

\bibitem[{{Soltan}(1982)}]{Soltan1982}
{Soltan}, A. 1982, \mnras, 200, 115

\bibitem[{{Springel}(2010)}]{Springel2010}
{Springel}, V. 2010, \araa, 48, 391

\bibitem[{{Springel} \&
  {Hernquist}(2003{\natexlab{a}})}]{SpringelHernquist2003}
{Springel}, V. \& {Hernquist}, L. 2003{\natexlab{a}}, \mnras, 339, 289

\bibitem[{{Springel} \& {Hernquist}(2003{\natexlab{b}})}]{Springel2003}
---. 2003{\natexlab{b}}, \mnras, 339, 289

\bibitem[{{Springel} {et~al.}(2001){Springel}, {Yoshida}, \& {White}}]{gadget}
{Springel}, V., {Yoshida}, N., \& {White}, S.~D.~M. 2001, New Astronomy, 6, 79

\bibitem[{{Springel} {et~al.}(2005)}]{Springel2005}
{Springel}, V. {et~al.} 2005, Nature, 435, 629

\bibitem[{{Thielemann} {et~al.}(2003){Thielemann}, {Argast}, {Brachwitz},
  {Hix}, {H{\"o}flich}, {Liebend{\"o}rfer}, {Martinez-Pinedo}, {Mezzacappa},
  {Panov}, \& {Rauscher}}]{Thielemann2003}
{Thielemann}, F.-K., {Argast}, D., {Brachwitz}, F., {Hix}, W.~R.,
  {H{\"o}flich}, P., {Liebend{\"o}rfer}, M., {Martinez-Pinedo}, G.,
  {Mezzacappa}, A., {Panov}, I., \& {Rauscher}, T. 2003, Nuclear Physics A,
  718, 139

\bibitem[{{Thorne}(1974)}]{Thorne1974}
{Thorne}, K.~S. 1974, ApJ, 191, 507

\bibitem[{{Tornatore} {et~al.}(2007){Tornatore}, {Borgani}, {Dolag}, \&
  {Matteucci}}]{Tornatore2007}
{Tornatore}, L., {Borgani}, S., {Dolag}, K., \& {Matteucci}, F. 2007, \mnras,
  382, 1050

\bibitem[{{Tornatore} {et~al.}(2004){Tornatore}, {Borgani}, {Matteucci},
  {Recchi}, \& {Tozzi}}]{Tornatore2004}
{Tornatore}, L., {Borgani}, S., {Matteucci}, F., {Recchi}, S., \& {Tozzi}, P.
  2004, \mnras, 349, L19

\bibitem[{{Tremaine} {et~al.}(1994){Tremaine}, {Richstone}, {Byun}, {Dressler},
  {Faber}, {Grillmair}, {Kormendy}, \& {Lauer}}]{Tremaine_et_al_1994}
{Tremaine}, S., {Richstone}, D.~O., {Byun}, Y., {Dressler}, A., {Faber}, S.~M.,
  {Grillmair}, C., {Kormendy}, J., \& {Lauer}, T.~R. 1994, \aj, 107, 634

\bibitem[{{van den Hoek} \& {Groenewegen}(1997)}]{vandenHoekGroenewegen1997}
{van den Hoek}, L.~B. \& {Groenewegen}, M.~A.~T. 1997, \aaps, 123, 305

\bibitem[{{Verdoes Kleijn} \& {de Zeeuw}(2005)}]{Verdoes2005}
{Verdoes Kleijn}, G.~A. \& {de Zeeuw}, P.~T. 2005, \aap, 435, 43

\bibitem[{{Wada} {et~al.}(2009){Wada}, {Papadopoulos}, \& {Spaans}}]{wada2009}
{Wada}, K., {Papadopoulos}, P.~P., \& {Spaans}, M. 2009, \apj, 702, 63

\bibitem[{{Wilson} \& {Colbert}(1995)}]{WilsonColbert1995}
{Wilson}, A.~S. \& {Colbert}, E.~J.~M. 1995, \apj, 438, 62

\bibitem[{{Woosley} \& {Weaver}(1995)}]{WW1995}
{Woosley}, S.~E. \& {Weaver}, T.~A. 1995, \apjs, 101, 181

\bibitem[{{Yu} \& {Tremaine}(2002)}]{YuTremaine2002}
{Yu}, Q. \& {Tremaine}, S. 2002, {MNRAS}, 335, 965

\end{thebibliography}

\label{lastpage}
\end{document}